\documentclass[10pt,twocolumn,english]{article}

\usepackage[T1]{fontenc}
\usepackage[latin9]{inputenc}
\usepackage[a4paper]{geometry}
\usepackage{color}
\usepackage{babel}

\usepackage{varioref}
\usepackage{amsmath}
\usepackage{graphicx}
\usepackage{amssymb}
\usepackage[unicode=true, 
 bookmarks=true,bookmarksnumbered=false,bookmarksopen=false,
 breaklinks=false,pdfborder={0 0 1},backref=false,colorlinks=false]
 {hyperref}
\hypersetup{
 pdfauthor={Philippe Maincon}}

\providecommand{\tabularnewline}{\\}

\begin{document}

\title{\noindent A Wiener-Laguerre model of VIV forces given recent cylinder
velocities}

\date{\noindent \today}

\author{\noindent Philippe Mainçon}

\maketitle

\section*{Abstract}

Slender structures immersed in a cross flow can experience vibrations
induced by vortex shedding (VIV), which cause fatigue damage and other
problems. VIV models in engineering use today tend to operate in the
frequency domain. A time domain model would allow to capture the chaotic
nature of VIV and to model interactions with other loads and non-linearities.
Such a model was developed in the present work: for each cross section,
recent velocity history is compressed using Laguerre polynomials.
The compressed information is used to enter an interpolation function
to predict the instantaneous force, allowing to step the dynamic analysis.
An offshore riser was modeled in this way: Some analyses provided
an unusually fine level of realism, while in other analyses, the riser
fell into an unphysical pattern of vibration. It is concluded that
the concept is promissing, yet that more work is needed to understand
orbit stability and related issues, in order to further progress towards
an engineering tool.

\section{\noindent Introduction}

Vortex induced vibration (VIV) is a vibration of an elastic structure
that occurs when a fluid flowing around the structure sheds vortices
at near-regular intervals, locked with the structure's own vibration.
VIV is a major concern in the offshore oil industry in particular,
where marine currents can cause slender structures like pipelines,
risers, umbilicals and cables to vibrate, inducing fatigue damage.
VIV is a hard problem because on one hand full hydrodynamic computations
of vortex sheddings from structures are as yet impractical, and on
the oher hand, it is challenging to simplify a strongly non-linear
dynamic system. Semi empirical VIV models provide the state of the
art of VIV engineering. They work by predicting added mass and excitation
coefficients on the basis of reduced frequency and amplitude of vibration,
and seek one or several oscillation modes that satisfy equilibrium.

\noindent Compared to such semi empirical VIV models, in the long
term an efficient time-domain VIV model would open new possibilities
:
\begin{enumerate}
\item \noindent Study of VIV on non-linear structures, for example studying
the damping effect of seafloor interaction in a steel riser, or using
a hysteretic cross section model for VIV on flexible pipes.
\item \noindent Accounting for VIV caused by unsteady water flows, in particular
by waves or vessel motions.
\item \noindent Accounting for the increase in drag at wave frequency due
to VIV.
\item \noindent Accounting for the superposition of wave-frequency and VIV-frequency
stresses in fatigue analysis.
\item Accounting for the asymmetry of oscillation patterns in the vicinity
of, for example, a seafloor.
\end{enumerate}
\noindent The objective of the work reported here is to demonstrate
the viability of a local, deterministic, time-domain force model for
VIV on slender bodies with cylindric cross sections. This force model
is used at each Gauss point of the dynamic finite element (FE) model
of a slender structure subject to external steady or unsteady water
currents, during a time domain analysis (e.g. Newmark-$\beta$ time
integration with Newton-Raphson iteration). Hence the FE model resembles
that commonly used in a slender structure analysis, with degrees of
freedom for the structure, and none for the surrounding fluid. In
other words, the proposed model takes the place usually held in software
by the Morison model for wave induced loads.

The litterature describes a few time-domain models of VIV that, like
the present model, do not explicitely model the wake flow. In \cite{lie95},
at any step and point along a cable, the recent velocity history is
approximated by a harmonic function, which is then used to enter charts
that predict excitation and added mass coefficients as a function
of reduced amplitude and frequency. A model which could be described
in the same way, but differs in several details was developped by
\cite{finn99}. More recently, the hydrodynamic force has been described
as the response on a non-linear single degree of freedom van der Pol
oscilator \cite{facchinetti03,mathelin05,violette07,violette09}.
The models enumerated here only deal with cross-flow vibration.

The present model differs from the above ones in that it treats in-line
and cross flow vibrations jointly, does not use a harmonic simplification
of the motion or forces, and does not reduce the wake response to
a single degree of freedom system.

\section{\noindent Model outline\label{sec:Model-outline}}

\subsection{\noindent Postulate\label{sub:Postulate}}

\noindent The present work hinges on the following postulate. \textit{The
force exerted by the surrounding fluid on a section of the slender
structure, is completely determined by the recent histories at that
section of the velocities of the structure and the undisturbed fluid.}
Several points in this sentence are worthy of discussion.

\noindent The {}``\textbf{force}'' includes the components usually
distributed into added mass, excitation forces, drag, lift etc...
. 

That the force\emph{ {}``}\textbf{\textit{\emph{at section of the
slender structure}}}\textit{\emph{'' is determined by the history
{}``}}\textbf{\textit{\emph{at that section}}}\textit{\emph{'' implies
a {}``strip theory'' in which it is excluded that motions of the
structure at a point A cause disturbances in the fluid that affects
the force at point B away from A. }}In other words, it is assumed
that there is no significant transmission of information in the axial
direction within the water (as opposed to within the slender structure).
This would be proved wrong if it turned out that unstable phenomena,
like boundary layer shedding, although transmitting little energy
along the structure, transmits information that steers how local hydrodynamic
energy is channeled at a given point along the structure. 

\noindent That the force should be {}``\textbf{completely determined}''
implies that the behavior of the structure is deterministic. This
does not contradict the observation of hysteretic response of short
cylinders mounted on elastic support. Uniqueness of forces given a
position does not imply uniqueness of static equilibrium. Neither
does {}``completely determined'' contradict the observation of irregular
and unpredictable responses to VIV: non-linear dynamic systems can
have a chaotic behavior. Still, complete determinism is provably wrong,
since a short vertical cylinder dragged at uniform speed through water
will experience oscillating lift forces. At any given moment, there
is nothing in the history of (constant) velocity that allows to predict
whether the lift is left or right. So the present work is based on
the bet that ignoring such {}``bifurcations'' still leaves us with
a useful model.

\noindent {}``\textbf{History}'' here relates to causality. The
force on the structure does not depend of future motion of the structure.
By contrast, frequency-domain models do not make an explicit distinction
here, typicaly requiring a steady state vibration. This can also be
contrasted to Morison's equation which predicts forces on a cylinder
from \emph{instantaneous} relative velocities and accelerations. 

{}``\textbf{Recent}'' can be defined as anything between the present
time and a few times $t_{w}$, where the value of $t_{w}$ still is
an object of debate. $t_{w}$ is likely to be case dependent. Current
will transport (convect) away vortices so that they quickly loose
significance, so that $t_{w}$ should be of the order of $D/U$ where
$D$ is the cross section diameter and $U$ the current velocity.
In contrast, if the cylinder is oscillating in still water, it will
be traveling in its own wake, and $t_{w}$ should be related to the
rate of diffusion and/or viscous dissipation of vortices, which is
likely to result in much higher values of $t_{w}$. Tests on periodic
forced motion of short cylinders sometimes show a slow drift of the
forces (over as many as ten periods). In contrast, force decay tests
for a cylinder stopped after oscilliations at zero mean velocity,
point towards a fraction of a period. In the present work, the idea
is to chose a {}``universal'' value of $t_{w}$ for the system,
after adequate scaling (cf. Section \ref{sec:Scaling}).

\noindent The {}``\textbf{velocities}'' are what count. Accelerations
would not do because for example, zero acceleration can correspond
to different speeds and hence different forces. On the other hand,
the force on a cylinder will not be affected by a uniform translation
of its whole trajectory, so a history of positions contains irrelevant
information.

\noindent In the remained of this text, the word {}``trajectory''
will be given a very specific meaning. The trajectory is defined as
the \emph{recent history of the velocity vector of the cylinder relative
to the undisturbed surrounding fluid}.

\subsection{\noindent Restrictions}

\noindent In the present phase of research, the following restrictions
are introduced, in order to achieve some simplification of the task.
The outer cross section of the slender structure is assumed perfectly
circular and smooth. The surrounding fluid is assumed to be infinite,
excluding the presence of sea floor, free surface or neighbouring
risers. Only fluid flows perpendicular to the cylinder at any point
are considered.

\subsection{\noindent Input and output}

\noindent As stated earlier, the VIV model being developed herel replaces
the Morison model for wave induced loads. The VIV model is called
at each step and iteration, and at each Gauss point or node of each
element. 

\noindent The model is to receive as input:
\begin{enumerate}
\item \noindent the diameter of the cylinder.
\item \noindent the instantaneous velocity of the cross section relative
to the undisturbed fluid. 
\item \noindent the instantaneous velocity and acceleration of the local
undisturbed fluid, in a Galilean reference system.
\end{enumerate}
\noindent The model uses velocity information stored from previous
steps. On this basis, the model produces as output:
\begin{enumerate}
\item \noindent the vector of hydrodynamic forces per unit length, acting
on the cylinder.
\item \noindent the matrix containing the derivative of the above with respect
to instantaneous values of the cylinder's behaviour.
\end{enumerate}
\noindent Gauss integration is then used to compute a consistent load
vector and derivative matrix for each element. Note that these element
matrices are likely to vary significantly over each VIV oscillation
{}``period'' - in contrast to added mass or damping matrices, deemed
to be constant over a long time in semi-empirical VIV models. The
connection of the force model to the finite element analysis is discussed
in Section \ref{sec:Dynamic-analysis}.

\subsection{\noindent \label{sub:Algorithmic-steps}Algorithmic steps}

\noindent Only the local VIV model is described here, not the whole
FE analysis.
\begin{enumerate}
\item \noindent The relative velocity of the cylinder relative to water
(thereafter: {}``velocity'') is computed.
\item \noindent \label{enu:scaling1}The velocity is scaled (Reynolds scaling)
to that the cylinder diameter is the unit of distance (Section \ref{sec:Scaling}).
\item \noindent \label{enu:The-recent-history} The trajectory (again: the
recent histories of both $x$ and $y$ components of velocity) is
compressed to a small number of {}``Laguerre coefficients''. This
compression is such that it provides accurate information over the
recent past and increasingly coarse information for more distant past
(Section \ref{sec:Tadpole}).
\item \noindent \label{enu:perceptron}The Laguerre coefficients are used
to enter an interpolation function (a feed-forward neural network
with some specifically tailored properties) which returns $x$ and
$y$ components of hydrodynamic force (Section \ref{sec:perceptron}).
The fitting of the interpolation function is discussed in Section
\ref{sub:Training-1}.
\item \noindent \label{enu:scaling2}The force is scaled back to the relevant
diameter (Section \ref{sec:Scaling})
\item \noindent \label{enu:The-Froude-Krylov-forces,}The Froude-Krylov
forces, which depend on the acceleration of the undisturbed flow,
are added (Section \ref{sub:Frode-Krylov-forces})
\end{enumerate}
The identification of non-linear systems using a bank of orthogonal
filters (including Laguerre filter) to generate multiple signals from
a single one, and then using the multiple signals to enter a non-linear,
memory-less function, was introduced by Norbert Wiener \cite{wiener58}.
In the present work, a base of Laguerre \emph{polynomials} is used,
in contrast to Laguerre \emph{functions }introduced by Wiener. While
Wiener apparently did not use neural networks as non-linear functions
(but for example Hermite polynomials), neural networks in Wiener models
have been studied for some time \cite{chen92}. In the present work,
Laguerre filtering is presented without making use of the vocabulary
of cybernetics. In particular, the $z$-transform is not introduced
here.

\section{\noindent Ancillary transformations\label{sec:Ancillary-transformations}}

\subsection{\noindent Froude-Krylov forces\label{sub:Frode-Krylov-forces}}

\noindent This section gives the justification for point \ref{enu:The-Froude-Krylov-forces,}
of Section \ref{sub:Algorithmic-steps}. If the undisturbed fluid
in which the cylinder is plunged is accelerating (because of surface
waves, for example), then it is natural to introduce two reference
systems: $\mathfrak{G}$ is a Galilean reference system, for example
fixed relative to the sea floor, $\mathfrak{A}$ is an accelerated
reference system, locally following the undisturbed flow. Transforming
the equations of equilibrium from $\mathfrak{G}$ (in which we carry
out FEM analysis) to $\mathfrak{A}$ (for which we have experimental
data, in water that is no accelerated) requires the addition of inertia
forces. 

\noindent The inertial forces create a uniform pressure gradient that
was not present in the laboratory test. The effect of a pressure gradient
on a submerged body is variously referred to as {}``Archimedes forces''
when the pressure gradient results from the acceleration of gravity,
or as {}``Froude-Krylov forces'' when the pressure gradient is due
to fluid acceleration in surface waves. As familiar, the integral
of the pressure over the wet surface is transformed into a volume
integral \cite{faltinsen90}.

\noindent It is assumed that this pressure gradient does not affect
the turbulent flow, so that the pressure gradient can simply be added
to the pressures resulting from turbulence. This seems reasonable
enough for incompressible flows, and indeed when it comes to Archimedes
forces, the submerged weight of a cylinder is routinely subtracted
to laboratory measurements and the relevant correction added again
in FEM analysis - even though the Archimedes forces in the laboratory
do not necessarily scale with those in the analysis. Further there
is no experimental indication that a horizontal and vertical cylinder,
all other conditions being equal, experience different forces. 

\noindent To conclude, the hydrodynamic force acting on the cylinder
at a given instant is the sum of two terms:
\begin{enumerate}
\item \noindent \label{enu:A-force-that}A force that is a function of only
the cylinder diameter and the recent history of the velocity of the
cylinder relative to the undisturbed, steady water flow.
\item \noindent Froude-Krylov forces.
\end{enumerate}
\noindent All computations in Sections \ref{sec:Tadpole} and \ref{sec:perceptron}
deal only with the first of the above two terms.

\subsection{\noindent \label{sec:Scaling}Scaling}

\noindent This section details how points \ref{enu:scaling1} and
\ref{enu:scaling2} of Section \ref{sub:Algorithmic-steps} are implemented.
In order to reduce the amount of experimental data necessary to create
the interpolation function used in point \ref{enu:perceptron}, one
must take advantage of scale similarities. To that effect, all data
used to either train or query the database is scaled. Correspondingly,
all forces returned by the database are scaled back. 

VIV forces are assumed to be uniquely defined by fluid density $\rho$,
kinematic viscosity $\nu$, cylinder diameter $D$ and the motion.
Hence, in order to create a database that is be entered with scaled
velocities, we wish all experimental data to be scaled to fixed reference
values $\rho_{o}$, $\nu_{o}$ and $D_{o}$. By expressing the units
of these quantities, one gets three equations on $\lambda_{m}$, $\lambda_{s}$
and $\lambda_{kg}$, which are the scaling factors for the basic units
of distance, time and mass. Solving the system yields\begin{eqnarray}
\lambda_{m} & = & D_{o}\frac{1}{D}\label{eq:scale_distance}\\
\lambda_{s} & = & \frac{D_{o}^{2}}{\nu_{o}}\frac{\nu}{D^{2}}\label{eq:scale_time}\\
\lambda_{kg} & = & \rho_{o}D_{o}^{3}\frac{1}{\rho D^{3}}\label{eq:scale_weight}\end{eqnarray}
Once the scaling of basic units is known, the scaling of any derived
quantities e.g. velocities, accelerations and forces per unit length
can be expressed:\begin{eqnarray}
\lambda_{ms^{-1}} & = & \frac{\nu_{o}}{D_{o}}\frac{D}{\nu}\label{eq:scale_velocity}\\
\lambda_{ms^{-2}} & = & \frac{\nu_{o}^{2}}{D_{o}^{3}}\frac{D^{3}}{\nu^{2}}\label{eq:scale_acceleration}\\
\lambda_{Nm^{-1}} & = & \frac{\rho_{o}\nu_{o}^{2}}{D_{o}}\frac{D}{\rho\nu^{2}}\label{eq:scale_force}\end{eqnarray}
Note that since scaling is applied consistently to all derived quantities,
all non-dimensional numbers based on combinations of distance, time
and mass (including Reynolds and Froude numbers) is conserved. However,
any dimensional quantity with units different from those of $\rho$,
$\nu$ and $D$ is scaled to values that depend of $\rho$, $\nu$
and $D$. In particular, Equation \ref{eq:scale_acceleration} shows
that all accelerations, including the acceleration of gravity $g$
are scaled with a factor proportional to $D^{3}/\nu^{2}$. So while
the scaling used here may conserve Froude's number, it does not allow
to build a database of forces related to surface wave effects, because
the database does not refer to a constant value $g_{o}$.

The choice of $\rho_{o}$, $\nu_{o}$ and $D_{o}$ is arbitrary, and
in this work, all are set to the value 1. $D_{o}=1\,[m]$ implies
that scaled displacements can be considered to have {}``1 diameter''
as unit. $D_{o}=1\,[m]$ and $\nu_{o}=1\,[m^{2}/s]$ together imply
that scaled velocities are expressed as Reynolds numbers since the
scaled velocity is calculated as $Dv/\nu$ where $v$is the velocity.

The Reynolds number is usualy computed using some velocity characteristic
of the system under study. In VIV science, the undisturbed velocity
of the current is used. By contrast, in this work, instantaneous local
values of the relative velocity vector is multiplied by $\frac{D}{\nu}$.
The scaled velocities thus obtained are a generalisation of the traditional
use of Reynolds number: Considering an immobile cylinder in a current,
the norm of its scaled relative velocity vector is equal to the traditional
Reynolds number. To prevent confusion of the present usage of Reynolds
number with the more particular classical one, yet emphasize the relation
between both, the expression {}``ilr-Reynolds'' (for {}``instant,
local, relative Reynolds'') will be used in this document.

\section{\noindent \label{sec:Tadpole}Characterization of trajectory}

\subsection{Foreword}

\noindent %
\begin{figure}
\includegraphics[bb=100bp 50bp 500bp 380bp,clip,width=8cm]{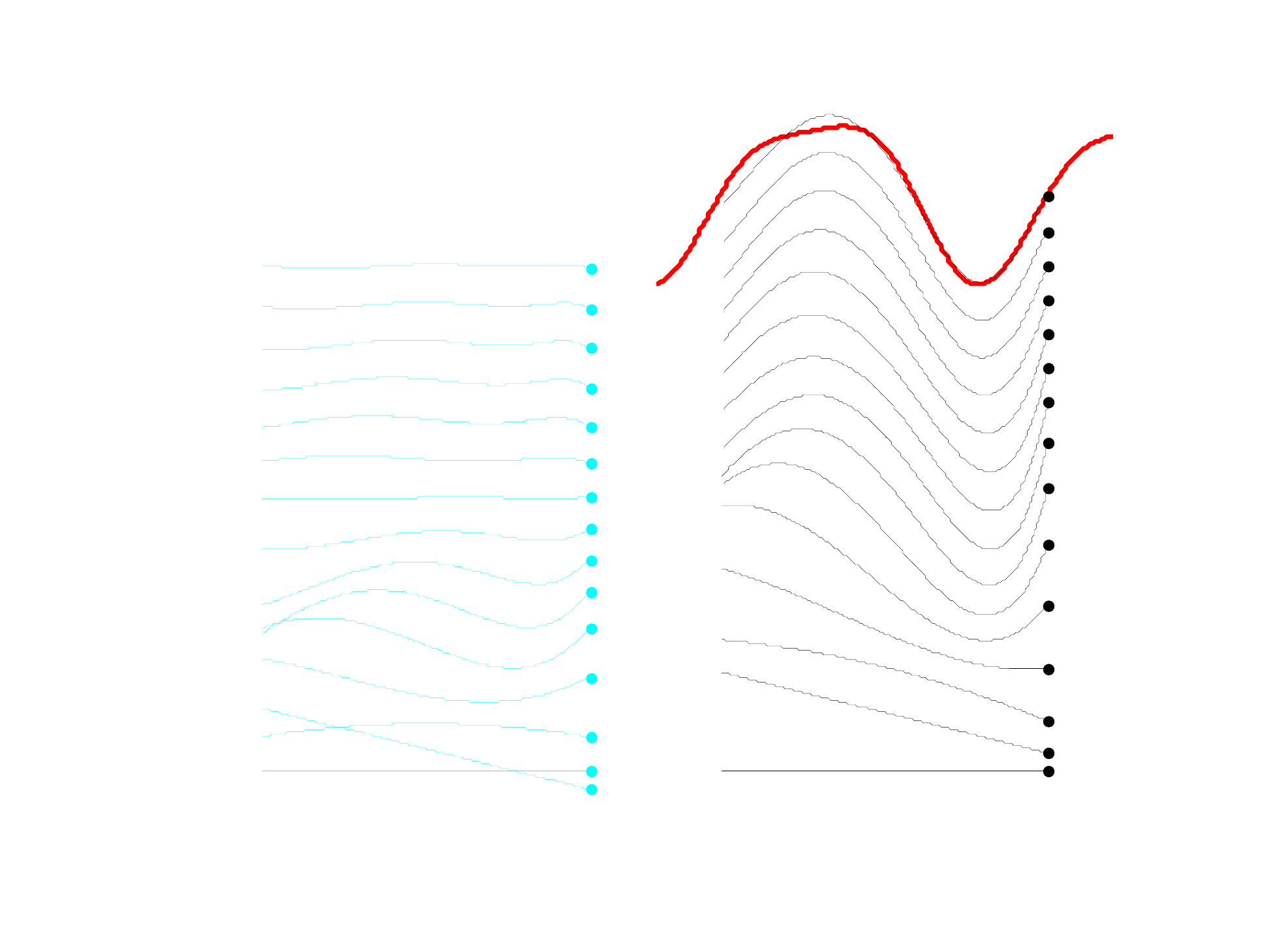}\caption{Weighted Laguerre polynomials (blue) are summed (black) to approximate
a trajectory (red). Vertical shifts were added for readability. The }
\label{Flo:LaguerrePrinciple}
\end{figure}

\noindent This section details how point \ref{enu:The-recent-history}
in Section \ref{sub:Algorithmic-steps} is to be implemented. The
objective is, for any given point in time, to distill a {}``summary''
of the recent history of the velocity of the cylinder relative to
the surrounding fluid (trajectory). Note that the history of each
component of the velocity vector is treated separately in this section
and that the procedure is applied to the \emph{scaled} trajectory.

\noindent The trajectory is approximated as a linear combination of
some adequate family of functions, and the coefficients in this linear
combination are the summary (Figure \ref{Flo:LaguerrePrinciple}.
The family of functions that is used here is the series of Laguerre
polynomials (Section\ref{sub:Definitions}). It is shown in Section
\ref{sub:Analysis-and-synthesis} that if the {}``Laguerre coefficients''
of the linear combination are obtained by integrating the product
of the trajectory by adequate {}``Laguerre dual'' functions, then
the difference between the approximating linear combination and the
real trajectory is small in the recent past and larger in the further
past. This justifies the choice of Laguerre polynomials: they allow
to summarise the trajectory in a way that represents recent velocities
very precisely, and older velocities in a coarser manner. It is assumed
that this corresponds to the information needed to obtain a good estimate
of the hydrodynamic force. 

Computing the integral of the product of Laguerre duals and trajectory
takes time. Luckily, one can show (Section \ref{sub:Differential-equation-for})
that the Laguerre coefficients are the solution of a differential
equation driven by the instant value of the velocity. To obtain results
that are independent of step size, this differential equation must
be carefully discretized in time (Section \ref{sub:Recursive-filter})
when summarizing experimental data.

\subsection{\noindent Definitions\label{sub:Definitions}}

\begin{figure}
\includegraphics[bb=10mm 0bp 407bp 305bp,clip,width=9cm]{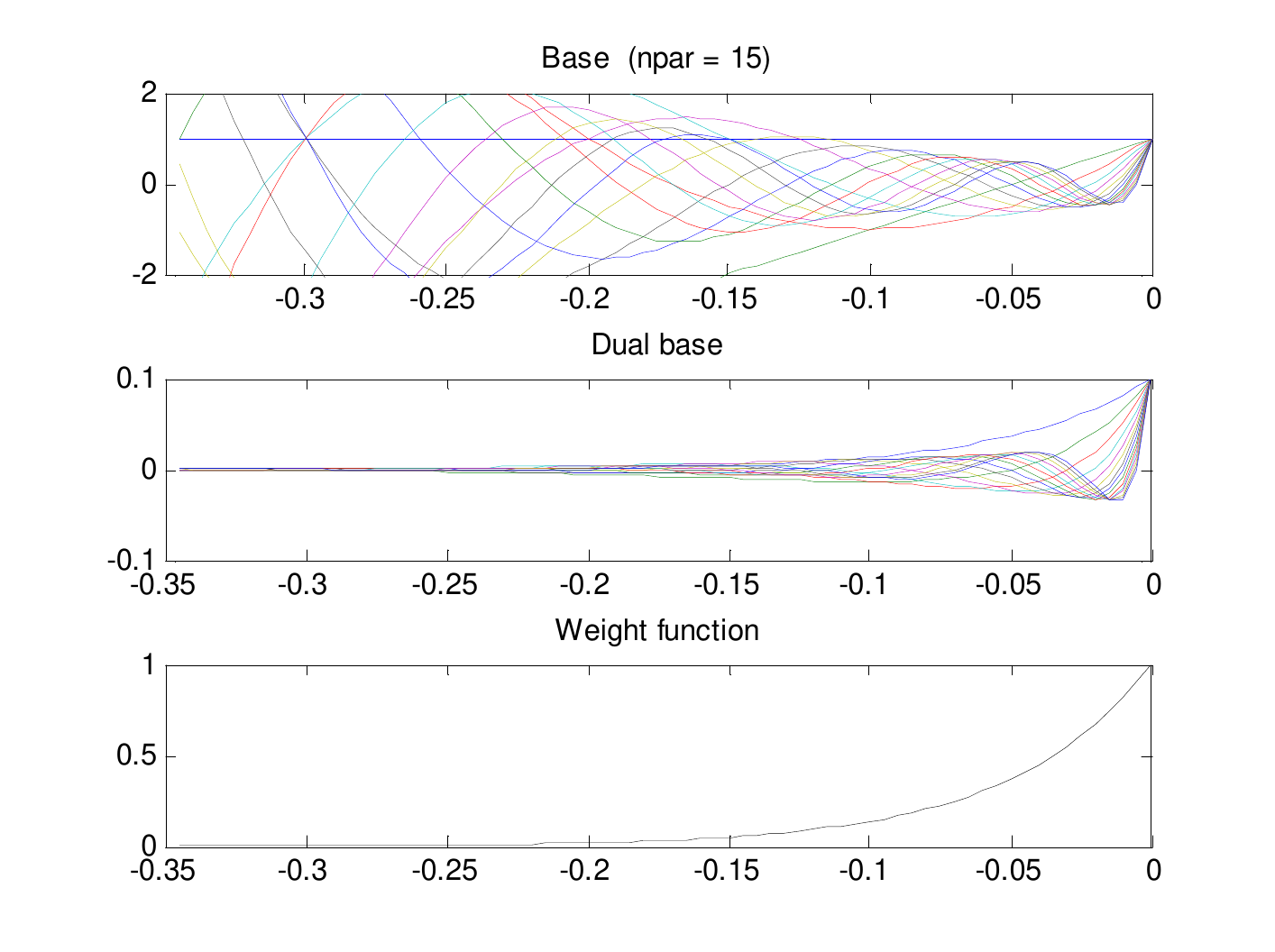}\caption{Laguerre polynomials (top), Laguerre duals (middle) and weight function
(bottom). Taking the convolution of a signal by the Laguerre duals
one obtains Laguerre coefficients. If one takes the linear combinations
of the Laguerre polynomials weighted by the coefficients, one gets
an approximation of the original signal, with a quality that decreses
towards the past in a way related to the weight function.}
\label{Flo:laguerre}
\end{figure}

\noindent The Laguerre polynomial\emph{ }(Figure \ref{Flo:laguerre},
top) of degree $i-1$ can be defined by its Rodrigues formula \cite{abramowitz64}\begin{equation}
\mathcal{L}_{i}\left(x\right)\equiv\frac{e^{x}}{i!}\frac{d^{i}}{dx^{i}}\left(x^{i}e^{-x}\right)\end{equation}
Laguerre polynomials verify the orthonormality property\begin{equation}
\int_{0}^{^{\infty}}\mathcal{L}_{i}\left(x\right)\mathcal{L}_{j}\left(x\right)e^{-x}dx=\delta_{ij}\label{eq:Laguerre-orth}\end{equation}

\noindent We seek to describe the recent trajectory with a precision
that is good for the immediate past, and decreasing for the further
past. To this end, we introduce a weight function which emphasizes
{}``recent past'' (Figure \ref{Flo:laguerre}, bottom) \begin{equation}
\mathcal{W}(t)\equiv\frac{e^{\frac{t}{t_{w}}}}{t_{w}}\quad t\in\mathbb{R^{\mathrm{-}}}\end{equation}
were the interpretation of $t_{w}$ has been discussed in Section
\ref{sub:Postulate}. Functions will now be noted as vectors (in a
Hilbert space), marked with overlined symbols. An indexed family of
functions will be noted as a matrix (symbols with double overline)
and so will a linear operator (a distributions of two variables).
We introduce the symmetric positive definite operator\begin{equation}
\overline{\overline{W}}(t_{1},t_{2})=\delta(t_{1},t_{2})\mathcal{W}(t_{1})\end{equation}
and a dot product in a suitable space of real valued functions\begin{equation}
\overline{f}^{T}\circ\overline{g}\equiv\intop_{-\infty}^{0}f(t)g(t)dt\end{equation}
with the canonical norm\begin{equation}
\left|\overline{f}\right|\equiv\sqrt{\overline{f}\circ\overline{f}}\end{equation}
Further we introduce the base\begin{equation}
\overline{\overline{L}}(t,i)\equiv\mathit{\mathcal{L}}_{i}(-t/t_{w})\quad t\in\mathbb{R^{\mathrm{-}}},\: i\in\left\{ 1,\ldots,n\right\} \end{equation}
Equation \ref{eq:Laguerre-orth} can be rewritten in matrix notation
as\begin{equation}
\overline{\overline{I}}=\overline{\overline{L}}^{T}\circ\overline{\overline{W}}\circ\overline{\overline{L}}\end{equation}
where $\bar{\bar{I}}$ is the $n\times n$ identity matrix. It is
useful to introduce the weighted norm or $w$-norm\textrm{\begin{equation}
\left|\overline{f}\right|_{w}\equiv\sqrt{\overline{f}\circ\overline{\overline{W}}\circ\overline{f}}\end{equation}
}

Note that since $\mathcal{W}(t)$ is of dimension $[1/s],$ \textrm{$\left|\overline{f}\right|_{w}$
}is of the same dimension as\textrm{$\overline{f}$}. So went taking
\textrm{$\overline{f}$} as a scaled velocity, \textrm{$\left|\overline{f}\right|_{w}$}
is an ilr-Reynolds number.

\subsection{\noindent Analysis and synthesis\label{sub:Analysis-and-synthesis}}

\noindent For a history $\overline{v}(t)$ of either the $x$ or $y$
component of the velocity, we seek the vector of {}``Laguerre coefficients''
$\overline{\tau}$ for the above base that minimize the weighted discretization
error\begin{eqnarray}
J & = & \frac{1}{2}\left|\overline{v}-\overline{\overline{L}}\cdot\overline{\tau}\right|_{w}^{2}\label{eq:target}\\
 & = & \frac{1}{2}\left(\overline{v}-\overline{\overline{L}}\cdot\overline{\tau}\right)^{T}\circ\overline{\overline{W}}\circ\left(\overline{v}-\overline{\overline{L}}\cdot\overline{\tau}\right)\end{eqnarray}
To this effect we require that the derivative be zero:\begin{equation}
\frac{\partial J}{\partial\overline{\tau}}=\overline{\overline{L}}^{T}\circ\overline{\overline{W}}\circ\overline{\overline{L}}\cdot\overline{\tau}-\overline{\overline{L}}^{T}\circ\overline{\overline{W}}\circ\overline{v}\end{equation}
which implies\begin{eqnarray}
\overline{\tau} & = & \left(\overline{\overline{L}}^{T}\circ\overline{\overline{W}}\circ\overline{\overline{L}}\right)^{-1}\cdot\overline{\overline{L}}^{T}\circ\overline{\overline{W}}\circ\overline{v}\\
 & = & \overline{\overline{L}}^{T}\circ\overline{\overline{W}}\circ\overline{v}\\
\overline{\tau} & = & \overline{\overline{D}}^{T}\circ\overline{v}\label{eq:tadpole}\end{eqnarray}
with\begin{equation}
\overline{\overline{D}}\equiv\overline{\overline{W}}\circ\overline{\overline{L}}\end{equation}
where the functions $\overline{\overline{D}}$ used for analysis consists
of the\emph{ Laguerre duals} (Figure \ref{Flo:laguerre}, middle.
Not to be confused with the Laguerre functions introduced in Equation
\ref{eq:laguerre-function}) \begin{eqnarray}
\overline{\overline{D}}(t,i) & = & \mathcal{D}_{i}(-t/t_{w})\\
 & = & \mathcal{L}_{i}(-t/t_{w})\frac{e^{\frac{t}{t_{w}}}}{t_{w}}\end{eqnarray}
Although \textrm{\begin{equation}
\overline{\overline{D}}^{T}\circ\overline{\overline{L}}=\overline{\overline{I}}\end{equation}
}the functions in $\overline{\overline{D}}$ and $\overline{\overline{L}}$
do not span the same space. Hence the appellation {}``dual base''
is abusive.

\subsection{\noindent Convergence\label{sub:Convergence} }

\noindent Laguerre functions, which can be defined as\begin{eqnarray}
\overline{\overline{F}}\left(t,i\right) & \equiv & \mathcal{F}_{i}(-t/t_{w})\\
 & \equiv & \mathcal{L}_{i}(-t/t_{w})\frac{e^{\frac{t}{2\: t_{w}}}}{2t_{w}}\label{eq:laguerre-function}\end{eqnarray}
or, in matrix notation\begin{equation}
\overline{\overline{F}}=\sqrt{\overline{\overline{W}}}\circ\overline{\overline{L}}\end{equation}
have been extensively studied. Series of Laguerre functions are known
to converge almost everywhere (under some conditions of continuity)
\cite{muckenhoupt70}. In matrix notation this result can be stated
as\begin{equation}
\lim_{n\rightarrow\infty}\left|\overline{\overline{F}}\cdot\overline{\overline{F}}^{T}\circ\overline{f}-\overline{f}\right|=0\end{equation}
This can be used to obtain a result on the convergence of series of
Laguerre polynomials. We introduce the change of variables\begin{equation}
\overline{f}=\sqrt{\overline{\overline{W}}}\circ\overline{g}\end{equation}
so that\begin{eqnarray}
\left|\overline{\overline{F}}\cdot\overline{\overline{F}}^{T}\circ\overline{f}-\overline{f}\right| & = & \!\!\!\!\left|\overline{\overline{F}}\cdot\overline{\overline{D}}^{T}\circ\overline{g}-\sqrt{\overline{\overline{W}}}\circ\overline{g}\right|\\
 & = & \!\!\!\!\left|\sqrt{\overline{\overline{W}}}\circ\left(\overline{\overline{L}}\cdot\overline{\overline{D}}^{T}\circ\overline{g}-\overline{g}\right)\right|\\
 & = & \!\!\!\!\left|\overline{\overline{L}}\cdot\overline{\overline{D}}^{T}\circ\overline{g}-\overline{g}\right|_{w}\end{eqnarray}
We hence have convergence in terms of the quality of approximation
that we are seeking, with emphasis on the recent past. Further, on
any finite (or {}``compact'') interval, convergence in the $w$-norm
is equivalent to convergence almost everywhere. So under some conditions
of continuity on $\overline{g}$, the series of Laguerre polynomials
obtained using $\overline{\overline{D}}$ as analysis functions converges
almost everywhere towards $\overline{g}$ in any finite interval.

\begin{figure}
\includegraphics[width=8cm]{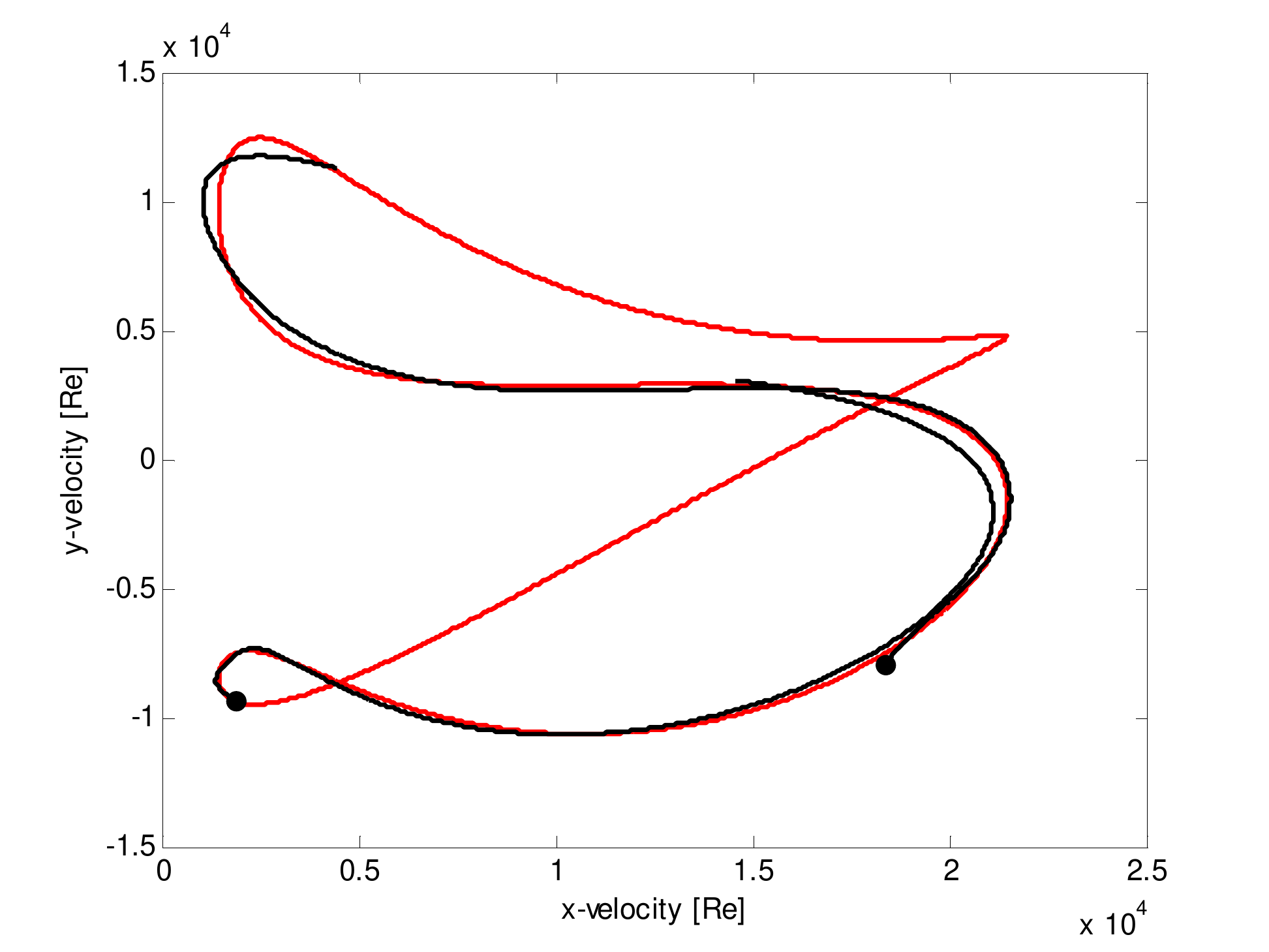}\caption{Example of Laguerre approximation for two components of a velocity
history (arbitrary scaling). The red dot marks the present time. The
red curve is the original cyclic signal and the black curves are Laguerre
approximations for two different instants}
\label{Flo:tadpole quality}
\end{figure}

Figure \ref{Flo:tadpole quality} illustrates how Laguerre coefficients
indeed provide a {}``summary'' of the trajectory

\noindent %

\subsection{\noindent Differential equation for Laguerre coefficients\label{sub:Differential-equation-for}}

\noindent In the finite element analysis, we need to update the Laguerre
coefficients at each iteration of each time step, for every Gauss
point of every node of the system. The explicit calculation of Equation
\ref{eq:tadpole} for every update is hence a CPU-time critical operation,
taking in the order of $n\times N$ floating point operations (flops),
where $n$ is the number of Laguerre polynomial used, and $N$ the
number of time steps that the dual functions take to decay to a negligible
value. Further, for each Gauss point, $2N$ velocity values need to
be stored, a severe memory requirement.

\noindent In the present Section and the next it is shown how the
computation of Equation \ref{eq:tadpole} can be carried out by a
recursive operation requiring no other storage than that of the Laguerre
coefficients and the last velocity values, and taking in the order
of $n\times n$ flops, which is advantageous because $ $$n\ll N$.
In this Section it is shown that $\overline{\tau}$ verifies a differential
equation driven by the history $\overline{v}$ of the velocity component.
In Section \ref{sub:Recursive-filter}, this differential equation
is solved time-step by time-step in a recursive update.

\noindent Equation \ref{eq:tadpole} can be rewritten without matrix
notation, and differentiated\begin{eqnarray}
\frac{\partial\tau_{i}}{\partial t} & = & \phantom{-\frac{1}{t_{w}}}\int_{0}^{+\infty}e^{-\theta}\mathcal{L}_{i}\left(\theta\right)\frac{\partial v}{\partial t}\left(t-t_{w}\theta\right)d\theta\\
 & = & -\frac{1}{t_{w}}\int_{0}^{+\infty}e^{-\theta}\mathcal{L}_{i}\left(\theta\right)\frac{\partial v}{\partial\theta}\left(t-t_{w}\theta\right)d\theta\end{eqnarray}
Multiplying by $t_{w}$ and integrating by parts yields\begin{multline}
t_{w}\frac{\partial\tau_{i}}{\partial t}=-\left[e^{-\theta}\mathcal{L}_{i}\left(\theta\right)v\left(t-t_{w}\theta\right)\right]\\
+\int_{0}^{+\infty}\left[-e^{-\theta}\mathcal{L}_{i}\left(\theta\right)+e^{-\theta}\frac{\partial}{\partial\theta}\mathcal{L}_{i}\left(\theta\right)\right]\\
v\left(t-t_{w}\theta\right)d\theta\end{multline}
A property of Laguerre polynomials is\begin{eqnarray}
\frac{\partial}{\partial\theta}\mathcal{L}_{i}\left(\theta\right) & = & -\mathcal{L}_{i-1}^{^{(1)}}\left(\theta\right)\\
 & = & -\sum_{j=0}^{^{i-1}}\mathcal{L}_{j}\left(\theta\right)\end{eqnarray}
where $\mathcal{L}_{i}^{^{(1)}}\left(\theta\right)$ is a generalized
Laguerre polynomial. Hence we can write\begin{multline}
t_{w}\frac{\partial\tau_{i}}{\partial t}=\mathcal{L}_{i}\left(0\right)v\left(t\right)-\tau_{i}\\
-\int_{0}^{+\infty}e^{-\theta}\sum_{j=1}^{^{i-1}}\mathcal{L}_{j}\left(\theta\right)v\left(t-t_{w}\theta\right)d\theta\end{multline}
\begin{eqnarray}
 & = & v\left(t\right)-\tau_{i}-\sum_{j=1}^{^{i-1}}\tau_{j}\\
 & = & v\left(t\right)-\sum_{j=1}^{^{i}}\tau_{j}\end{eqnarray}
which is of the form\begin{equation}
\frac{\partial\overline{\tau}}{\partial t}\left(t\right)=\overline{\overline{\mu}}\cdot\overline{\tau}\left(t\right)+\overline{n}\; v\left(t\right)\label{eq:dual-diffeq}\end{equation}
with\begin{equation}
\begin{cases}
\mu_{ij}=-\frac{1}{t_{w}} & \quad j\leq i\\
\phantom{\mu_{ij}}=0 & \quad j>i\end{cases}\end{equation}
\begin{equation}
n_{i}=\frac{1}{t_{w}}\end{equation}

\noindent Equation \ref{eq:dual-diffeq} shows that at any time $t$,
the rate of the Laguerre coefficients is fully defined by the Laguerre
coefficients and the velocity signal.

\subsection{Recursive filter\label{sub:Recursive-filter}}

The discrete integration of Equation \ref{eq:dual-diffeq} must be
done carefully, for two reasons. First it is important to obtain Laguerre
coefficients that are independent of the sampling rate used (as long
as the sampling rate is {}``adequate''). This is because the experimental
data on which the VIV model is based may come from experiments which,
after scaling, may have different sampling rates. Further, the numerical
analysis in which the VIV model is used may use yet another time step.
The choice of time step or sampling rate must not affect the way a
trajectory is characterized by Laguerre coefficient. 

The second reason for care in discrete integration is that we wish
to be able to create synthesized signals $\overline{\overline{L}}\cdot\overline{\tau}$
of good quality. Synthesized signal are neither used in the numerical
process of creating a force interpolation function (Section \ref{sec:perceptron})
or in the FEM use of the VIV model. However visualization is essential
to the process of research, both for fault diagnosis and quality control,
and to communicate an understanding of the method.

This discrete integration is only used in the analysis of experimental
data, to provide an input to the training of the {}``rotatron''
(Section \ref{sub:Training}). In dynamic analysis, the integration
of Equation \ref{eq:dual-diffeq} is done by means of the Newmark-$\beta$
method, as detailed in Section \ref{sec:Dynamic-analysis}.

Assume that velocity is sampled at regular intervals\begin{equation}
v_{j}=v\left(t_{0}+j\, dt\right)\end{equation}
We seek the values of the Laguerre coefficients at the same intervals\begin{equation}
\overline{\tau}_{j}=\overline{\tau}\left(t_{0}+j\, dt\right)\end{equation}

The vector $\overline{\tau}_{j}$ (the list of the coefficients for
all Laguerre polynomial, take at step $j$) must not be confused with
scalar $\tau_{i}$ (the coefficient for the Laguerre polynomial of
degree $i$). We choose $t_{0}$ such that $t_{0}+j\, dt=0$, and
we approximate $v$ by a function that is linear over the interval
$[0,dt]$. Equation \ref{eq:dual-diffeq} becomes\begin{equation}
\frac{\partial\overline{\tau}}{\partial t}\left(t\right)=\overline{\overline{\mu}}\cdot\overline{\tau}\left(t\right)+\overline{\alpha}+\overline{\beta}t\label{eq:approx-dual-diffeq}\end{equation}
with\begin{eqnarray}
\overline{\alpha} & = & \overline{n}\, v\left(0\right)\\
\overline{\beta} & = & \overline{n}\,\frac{v\left(dt\right)-v\left(0\right)}{dt}\end{eqnarray}
This new differential equation can be solved exactly: We seek a solution
of the form\begin{equation}
\overline{\tau}\left(t\right)=\exp\left(\overline{\overline{\mu}}\, t\right)\cdot\overline{a}+\overline{b}t+\overline{c}\label{eq:ansatz}\end{equation}
over the interval. Here $\exp\left(\overline{\overline{\mu}}\, t\right)$
stands for a matrix exponential. Replacing this expression into Equation
\ref{eq:approx-dual-diffeq}, noting that\begin{eqnarray}
\frac{\partial}{\partial t}\exp\left(\overline{\overline{\mu}}\, t\right) & = & \overline{\overline{\mu}}\cdot\exp\left(\overline{\overline{\mu}}\, t\right)\\
\exp\left(\overline{\overline{0}}\right) & = & \overline{\overline{I}}\end{eqnarray}
and identifying the constant and linear terms and enforcing the initial
value leads to \begin{eqnarray}
\overline{b} & = & -\overline{\overline{\mu}}^{-1}\cdot\overline{\beta}\\
\overline{c} & = & -\overline{\overline{\mu}}^{-2}\cdot\overline{\beta}-\overline{\overline{\mu}}^{-1}\cdot\overline{\alpha}\\
\overline{a} & = & \overline{\tau}\left(0\right)+\overline{\overline{\mu}}^{-2}\cdot\overline{\beta}+\overline{\overline{\mu}}^{-1}\cdot\overline{\alpha}\end{eqnarray}
Replacing these expressions in Equation \ref{eq:ansatz} at $t=dt$,
a tedious but straightforward computation yields the recursive filter\begin{equation}
\overline{\tau}_{j+1}=\overline{\overline{M}}\cdot\overline{\tau}_{j}+\overline{V}_{1}\cdot v_{j}+\overline{V}_{2}\cdot v_{j+1}\end{equation}
with\begin{eqnarray}
\overline{\overline{M}} & = & \exp\left(\overline{\overline{m}}\, dt\right)\\
\overline{\mu}_{1} & = & \overline{\overline{\mu}}^{-1}\cdot\overline{n}\\
\overline{\mu}_{2} & = & \overline{\overline{\mu}}^{-2}\cdot\overline{n}\,\frac{1}{dt}\\
\overline{V}_{1} & = & \overline{\overline{M}}\cdot\left(\overline{\mu}_{1}-\overline{\mu}_{2}\right)+\overline{\mu}_{2}\\
\overline{V}_{2} & = & \overline{\overline{M}}\cdot\overline{\mu}_{2}-\overline{\mu}_{1}-\overline{\mu}_{2}\end{eqnarray}

\section{\noindent \label{sec:perceptron}Force interpolation}

\subsection{Foreword}

\noindent This section details the implementation of point \ref{enu:perceptron}
in Section \ref{sub:Algorithmic-steps}. This section presents an
interpolation function which, given the Laguerre coefficients, predicts
the present value of the force vector. Polynomials were considered
initially, but is soon became clear that feed-forward {}``neural
networks'' provide a better class of functions to work with. The
reason for that is that the number of polynomial coefficients of degree
$d$ for a polynomial of $n$ variables is $n^{d}$, and high values
of $d$ must be expected to be necessary. By contrast, in a neural
network, non-linearity is introduced by {}``sigmoid'' or {}``threshold''
functions, and the coefficients are used to specify in which direction
non-linearity applies. Further, polynomials are infamous for their
propensity to oscillate.

\noindent The {}``rotatron'' presented here is based on the {}``perceptron''
\cite{rosenblatt58,rojas96}, a well studied architecture of neural
network which provides a flexible tool for the interpolation of scalar-valued
functions of a vector (Section \ref{sub:Perceptron}). The rotatron
takes advantage of certain symmetry properties of the physics at hand
(Section \ref{sub:Rotational-symmetries}).

In Section \ref{sec:Dynamic-analysis}, the rotatron is used to predict
\emph{scaled }forces based on the Laguerre coefficients for \emph{scaled}
trajectories.

\subsection{\noindent \label{sub:Perceptron}Perceptron}

\noindent %
\begin{figure}
\includegraphics[bb=150bp 140bp 650bp 950bp,clip,width=8cm]{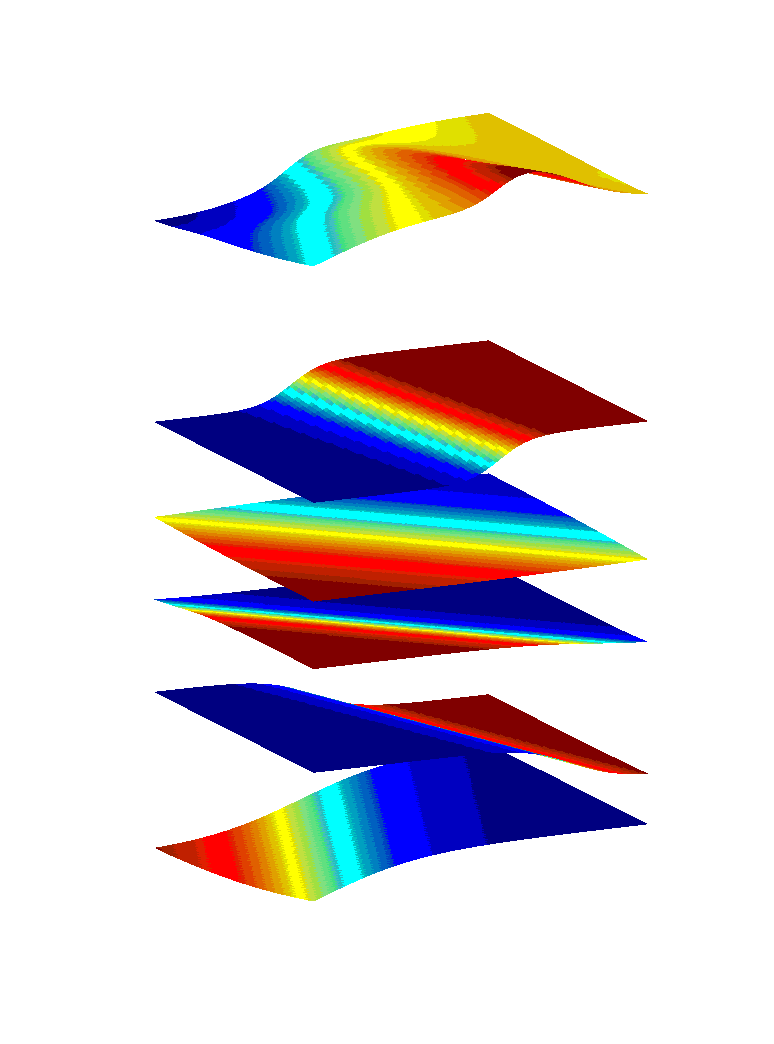}\caption{Perceptron: {}``arbitrary'' functions (top) can be represented as
a sum of sigmoid steps (bottom 5) of different orientation and steepness
$N_{jkl}$, shift $V_{j}$, and height $M_{ij}$ }
\label{Flo:perceptron}
\end{figure}

\noindent A perceptron \cite{rosenblatt58,rojas96} is a simple feed-forward
neural network, consisting of 3 layers. The input layer has $2n$
neurons where $n$ is number of Laguerre coefficients for each velocity
component and the factor 2 comes from the need to analyze in-line
and cross-flow speed histories together. The values of the input layer
neurons are set to the Laguerre coefficients for both velocity components.
The second layer has $nhid$ neurons, whose values are an affine function
of the values of the first layer, passed through a sigmoid function
like\begin{equation}
\sigma(x)=1-\frac{2}{e^{2x}+1}\label{eq:sigmoid}\end{equation}
Finally, the third layer gives the output of the perceptron, and its
values are an affine function of the values of the second layer. This
can be summarized as\begin{equation}
\hat{f_{i}}=M_{ij}\cdot\sigma\left(N_{jkl}\cdot\tau_{kl}+V_{j}\right)+U_{i}\label{eq:perceptron}\end{equation}
$M_{ij}$, $N_{jkl}$, $U_{i}$ and $V_{j}$ are the {}``weights''
or interpolation coefficients, that must be adjusted to fit the perceptron
to interpolate some given data. $\tau_{kl}$ are Laguerre coefficients
and $\hat{f_{i}}$ are predicted force components. $i$ is the index
of force direction ($x$ vs. $y$), $j$ the index of neuron in the
hidden layer, $k$ the index of velocity direction and $l$ index
of Laguerre coefficient.

Each output of the perceptron can be seen as a function, which is
a sum of sigmoid steps (Figure \ref{Flo:perceptron}).

\subsection{\noindent \label{sub:Rotational-symmetries}Symmetries}

\noindent %
\begin{figure}
\includegraphics[width=9cm]{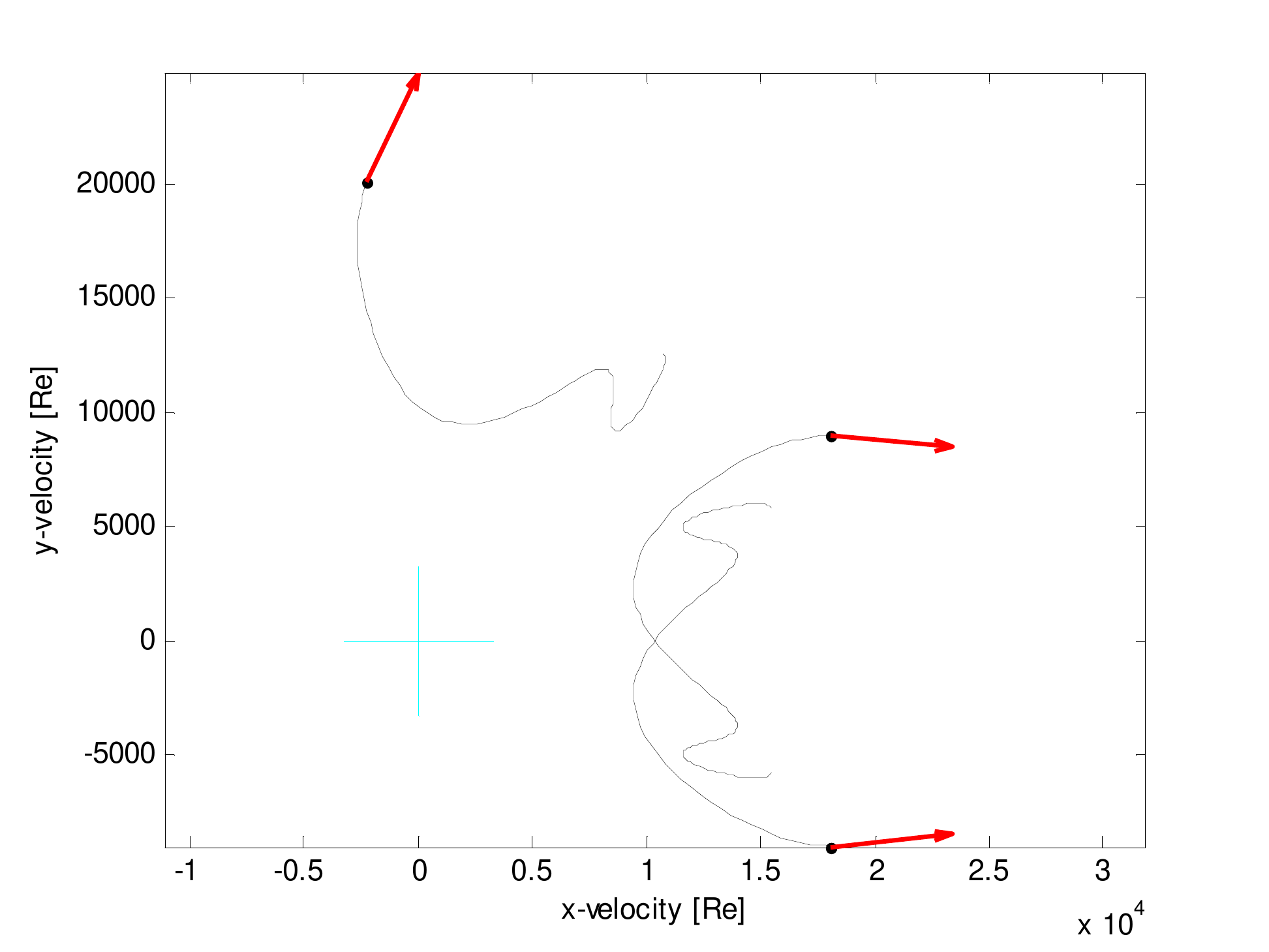}\caption{For circular cross section it is assumed that if two trajectories
can deduced from each other by rotation or mirroring, then the corresponding
forces are deduced from each other by the same operation}
\label{Flo:symmetries}
\end{figure}
The relation between trajectories (in the sense of history of the
velocity of the cylinder relative to the water) and forces can reasonably
be assumed to exhibit several symmetries (Figure \ref{Flo:symmetries}): 
\begin{description}
\item [{Rotational$\;$symmetry}] \noindent If a trajectory can be deduced
from the other by a rotation around the origin, then the resulting
forces are also deduced from each other by the same rotation. 
\item [{Mirror$\;$symmetry}] \noindent If a trajectory can be deduced
from the other by a mirroring around a line crossing the origin, then
the resulting forces are also deduced from each other by the same
mirroring. 
\end{description}
\noindent Rotational symmetry and mirror symmetry together, imply
directionality: If a trajectory is within a line crossing the origin,
then the resulting forces are within the same line. In particular
zero velocities must imply zero forces.

\noindent The symmetries imply that, once experimental data for a
trajectory has been obtained, there is no need to acquire data for
rotated or mirrored trajectories. However, if one was training a perceptron
to interpolate the data, the training set would need to include trajectories
\emph{and their rotates and mirrors}, with the correspondingly rotated
and mirrored forces. This would increase memory and CPU usage during
training, but also during use of the trained perceptron, because the
perceptron will need a larger number of hidden layer to interpolate
the training data.

\noindent Another approach is hence used in the present work: the
classic perceptron is replaced by a {}``rotatron'' (Section \ref{sub:Rotatron}).
It is designed so that, whatever the values of the weight coefficient,
a rotation or mirroring of the input trajectory results in the same
rotation or mirroring of the output force vector.

\subsection{\noindent \label{sub:Rotatron}Rotatron}

\begin{figure*}
\noindent \centering{}\includegraphics{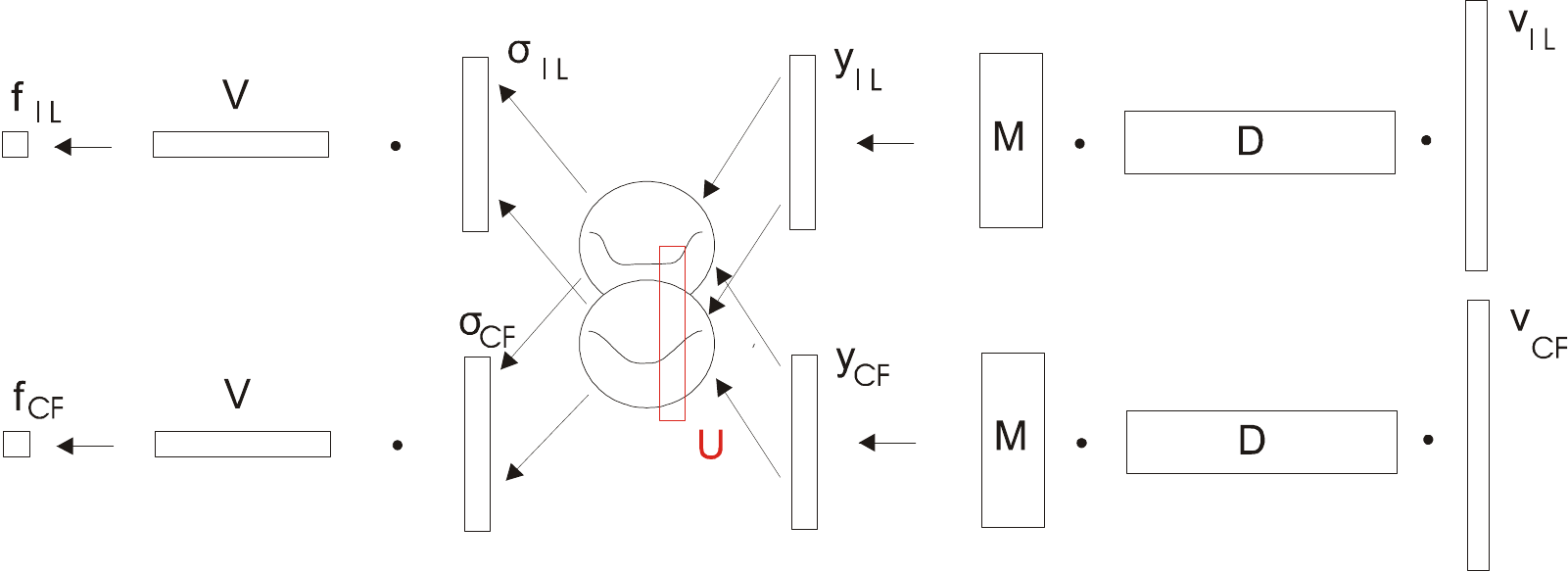}\caption{Laguerre analysis and rotatron transform velocity histories into a
hydrodynamic force. The matrix $D$ is the discrete form of the Laguerre
{}``duals'', which appear in Equation \ref{eq:tadpole}.}
\label{Flo:rotatron_graph}
\end{figure*}

\noindent A modified interpolation function (which will be refered
to as {}``rotatron'' in this text), which enforces the symmetries
discussed in Section \ref{sub:Rotational-symmetries}, is is defined
as 

\noindent \begin{equation}
\hat{f_{i}}=V_{k}\sigma_{ik}\label{eq:rot1}\end{equation}
with\begin{eqnarray}
\sigma_{ik} & = & \sigma_{i}\left(y_{[j]k}\right)\label{eq:dum}\\
 & = & \frac{y_{ik}}{|y_{[j]k}|\left(1+|y_{[j]k}|{}^{\alpha_{k}}\right)}\label{eq:sigmoid1}\\
|y_{[j]k}| & = & \sqrt{y_{1k}^{2}+y_{2k}^{2}}\label{eq:rot3a}\\
\alpha_{k} & = & -1-e^{-U_{k}}\label{eq:rot4}\\
y_{jk} & = & M_{kl}\,\tau_{jl}\label{eq:rot5}\end{eqnarray}
In the above, index $i$ and $j$ refer to direction, index $l$ to
the Laguerre polynomial and $k$ to the hidden layer. $V_{k}$, $U_{k}$
and $M_{kl}$ are tunable parameters. $\tau_{jl}$ are Laguerre coefficients,
given as input to the {}``rotatron''. See Appendix \ref{sec:Conventions-for-indexed}
for conventions on index notations and in particular for the syntax
$|y_{[j]k}|$. Note that Equations \ref{eq:tadpole}, \ref{eq:rot1}
and \ref{eq:rot5} operate linearly, identicaly and independently
on the terms related to the $x$ and $y$ directions, while Equation
\ref{eq:sigmoid1} involves a unit vector multiplied by a non linear
function of its norm. Figure \ref{Flo:rotatron_graph} illustrates
the flow of information, from right to left, from two vectors containing
the histories of the velocity components, to Laguerre coefficient,
that are then processed in the rotatron.%

\begin{figure}
\includegraphics[bb=9bp 5bp 560bp 420bp,width=9cm]{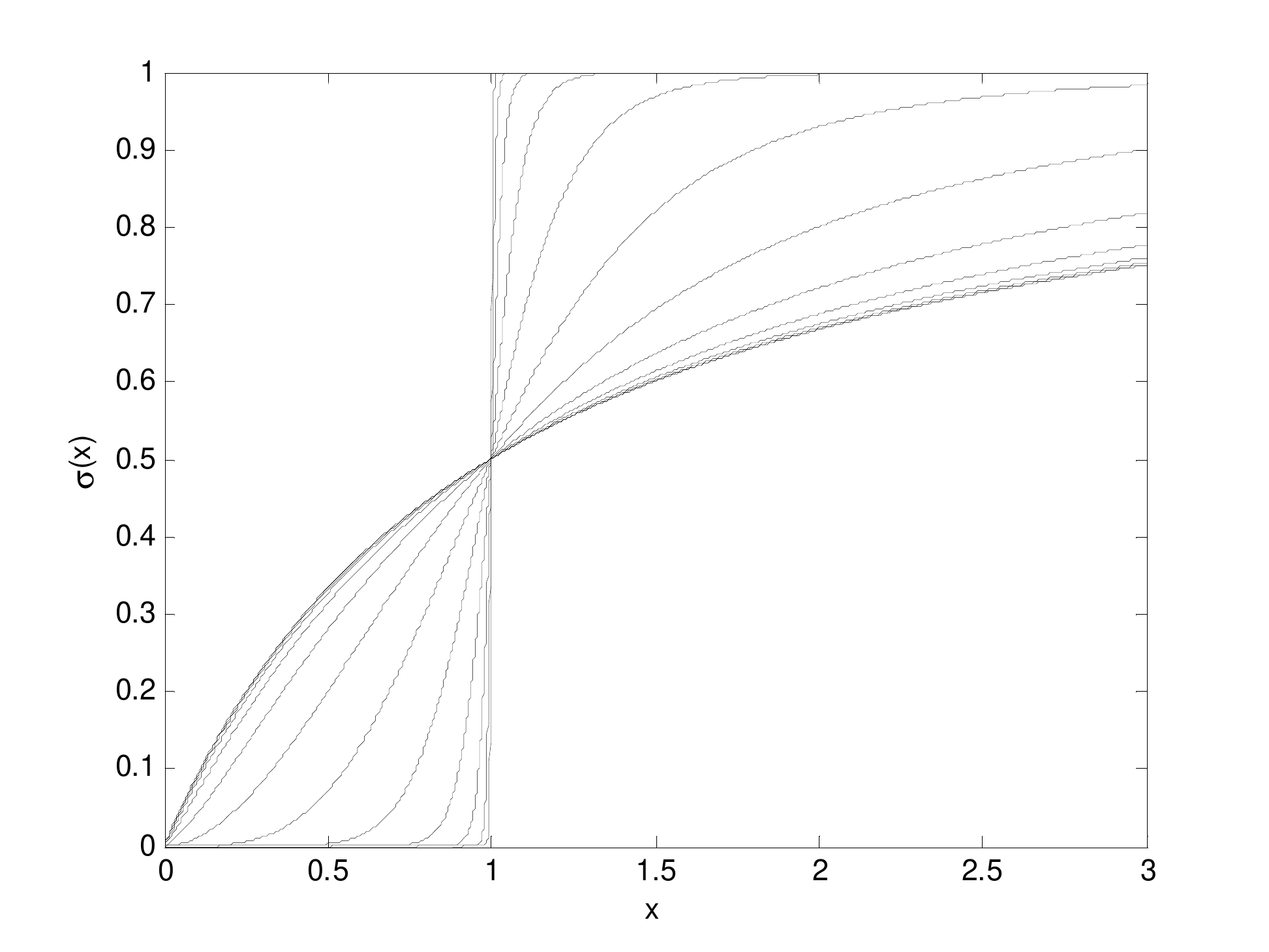}\caption{Log-logistic sigmoid functions}
\label{Flo:loglogi}
\end{figure}

\noindent The non-linear function appearing in Equation \ref{eq:sigmoid1}
is a sigmoid, whose abruptness is parametrized by $U_{k}$ (Figure
\ref{Flo:loglogi}). The sigmoid is shown in Figure \ref{Flo:loglogi}
for various values of the parameter $U_{k}$.%

\subsection{\noindent Training\label{sub:Training}}

\noindent {}``Training'' of a neural network refers to finding weight
coefficients $V_{k}$, $M_{kl}$ and, $U_{k}$ such that for any training
point number $m$, consisting of Laguerre coefficients $\tau_{jlm}$
and two force components $f_{im}$, the outputs \textrm{$\hat{f}_{im}$}
computed by the neural network are close to $f_{im}$.

\subsubsection{\noindent Regularization}

\noindent A common problem when training neural networks is {}``overspecialization''
\cite{sjoberg92}. In this situation, the neural network predicts
the training outputs with high accuracy but behaves wildly between
the training points. In contrast, what is implicitly sought is a smooth
response of the network to the input, even if this means an imperfect
fit to the training data.

\noindent Many strategies are described in the literature to address
this problem. One of them, which is adopted here, is regularization
\cite{sjoberg92}: the value of the weight parameters $V_{k}$, $M_{kl}$
and, $U_{k}$ are chosen by minimizing the cost function\begin{multline}
J\left(V_{[k]},U_{[k]},M_{[k]},f_{[im]},\tau_{[jlm]}\right)=\\
\frac{1}{2}\left(f_{im}-\hat{f}_{i}\left(\tau_{[jl]m}\right)\right)^{2}+\rho\frac{1}{2}\left(U_{k}^{2}+V_{k}^{2}+M_{kl}^{2}\right)\end{multline}
where $\rho$ is the regularization coefficient, an arbitrary input
to the training algorithm. High values of $\rho$ favor smoothness
of the response of the neural network against precision in reproducing
the training set.

\subsubsection{\noindent Conjugate gradient optimization}

\noindent $J$ is a function of a large number of weight coefficients,
and hence it is not practical to compute the Hessian of $J$, because
the Hessian is a full matrix. It also proves to be very costly to
even compute an approximation to it as done in the Levenberg-Marquardt
algorithm \cite{levenberg44,marquardt63}. On the other hand, the
Nelder-Mead {}``downhill simplex'' algorithm \cite{nelder65}, which
uses only the values of $J$, proved very slow in this case. Hence
a search method is chosen, that determines the search direction from
the gradient of $J$ \cite{moller93}. This is a conjugate gradient
method, in which the step length is found by deriving the gradient
in the direction of the search. In this method, the positive definiteness
of the (implicit) Hessian is forced by adding a scaled identity matrix
to it, a technique known as {}``trust region''. 

The conjugate gradient method proved far more efficient than the Levenberg-Marquardt
and Nelder-Mead methods for the present task.

\section{\noindent Metric\label{sec:Metric}}

\subsection{\noindent Euclidean metric and distance}

\noindent In order to describe the available data, it is useful to
define a distance between trajectories. This will allow to determine
to what extend the set of available data {}``fills'' the set of
all possible trajectories, or to detect zones of transition from one
hydrodynamic behavior to the other. Finally, this will help detecting
contradictions in the available data, arising from a variety of sources,
including hidden experimental variables, measurement uncertainties
or inadequate modeling in inverse methods and not least, the natural
variability of VIV forces.

\begin{figure}
\includegraphics[width=9cm]{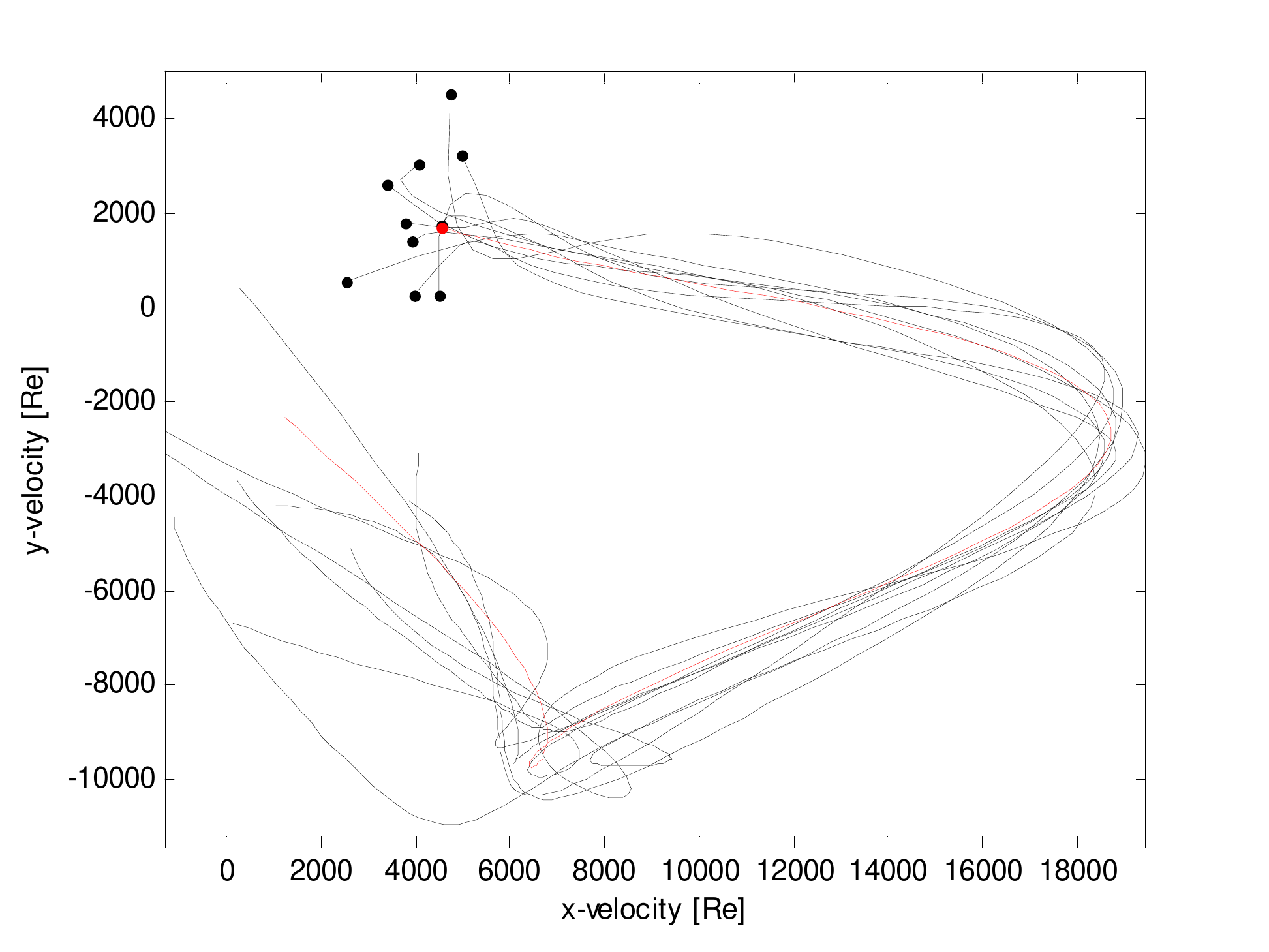}\caption{A set of neighboring trajectories according to Equation \ref{eq:euclid-dist}.
Typical distance between trajectories: $10^{3}\;[ilr\, Re]$.}
\label{Flo:MCtadpole}
\end{figure}

\noindent The $x$ and $y$ components of a trajectory are described
by a pair of functions:\begin{equation}
\overline{f}\equiv\left(\overline{f}_{x},\overline{f}_{y}\right)\end{equation}
We can define a scalar product between trajectories, that captures
any recent differences:\begin{equation}
\overline{f}^{T}\diamond\overline{g}\equiv\overline{f}_{x}^{T}\circ\overline{\overline{W}}\circ\overline{g}_{x}+\overline{f}_{y}^{T}\circ\overline{\overline{W}}\circ\overline{g}_{y}\end{equation}
\begin{equation}
=\intop_{-\infty}^{0}\frac{e^{\frac{t}{t_{w}}}}{t_{w}}\left(f_{x}(t)g_{x}(t)+f_{y}(t)g_{y}(t)\right)dt\end{equation}
By replacing $\overline{f}_{x}$, $\overline{f}_{y}$, $\overline{g}_{x}$
and $\overline{g}_{y}$ by their expression in terms of Laguerre polynomials
and their respective Laguerre coefficients $\overline{\tau}_{fx}$,
$\overline{\tau}_{fy}$, $\overline{\tau}_{gx}$ and $\overline{\tau}_{gy}$,
one finds that \begin{eqnarray}
\overline{f}^{T}\diamond\overline{g} & = & \overline{\tau}_{fx}^{T}\cdot\overline{\tau}_{gx}+\overline{\tau}_{fy}^{T}\cdot\overline{\tau}_{gy}\\
 & = & \overline{\tau}_{f}^{T}\cdot\overline{\tau}_{g}\end{eqnarray}
with\begin{equation}
\overline{\tau}_{f}\equiv\left[\begin{array}{c}
\overline{\tau}_{fx}\\
\overline{\tau}_{fy}\end{array}\right],\quad\overline{\tau}_{g}\equiv\left[\begin{array}{c}
\overline{\tau}_{gx}\\
\overline{\tau}_{gy}\end{array}\right]\end{equation}
The distance is defined from the scalar product in the usual manner:\begin{eqnarray}
\left|\overline{f}-\overline{g}\right| & \equiv & \sqrt{\left(\overline{f}-\overline{g}\right)^{T}\diamond\left(\overline{f}-\overline{g}\right)}\\
 & = & \left|\overline{\tau}_{f}-\overline{\tau}_{g}\right|\label{eq:euclid-dist}\end{eqnarray}
In other words, neighboring vectors of Laguerre coefficients describe
trajectories that are similar in the recent past. This is illustrated
by taking random samples of Laguerre coefficients around a given value
obtained from data analysis and plotting the synthesized trajectories
(Figure \ref{Flo:MCtadpole}).

\noindent %

\subsection{\noindent Rotatron-distance}

\noindent The above does not account for rotational and mirror symmetries.
We seek a distance for which the distance of a trajectory to its transforms
by rotation or mirroring is zero. Another distance is hence introduced:\begin{multline}
d\left(\overline{f},\overline{g}\right)\equiv\\
\min\left(\min_{R\in\mathfrak{R}}\left|\overline{f}-R\left(\overline{g}\right)\right|\;,\;\min_{S\in\mathfrak{S}}\left|\overline{f}-S\left(\overline{g}\right)\right|\right)\label{eq:rotatron-dist}\end{multline}
where $\mathfrak{R}$ is the set of all rotations of the trajectories
around the origin and $\mathfrak{S}$ the set of all mirroring of
trajectories around a line passing by the origin. Note that no norm
or scalar product associated to the distance $d$ is presented here
(The vector-space of trajectories, divided by the group of rotations
and mirrorings, is not a vector space). 

\noindent Because $\overline{f}_{x}$ is related to $\overline{\tau}_{fx}$
by the same linear relation that relates $\overline{f}_{y}$ to $\overline{\tau}_{fy}$,
linear combinations of $\overline{f}_{x}$ and $\overline{f}_{y}$
(including rotation and mirroring) are related to the same linear
combinations on $\overline{\tau}_{fx}$ and $\overline{\tau}_{fy}$.
By expressing the distance $\left|\overline{f}-R\left(\overline{g}\right)\right|$
as a function of the angle $\alpha$ of the rotation $R$, and then
differentiating with respect to $\alpha$, it can be shown that the
value of $\alpha$ that minimizes $\left|\overline{f}-R\left(\overline{g}\right)\right|$
is\begin{multline}
\alpha=\arctan\\
\left(\overline{\tau}_{fx}\cdot\overline{\tau}_{gy}-\overline{\tau}_{fy}\cdot\overline{\tau}_{gx}\;,\;\overline{\tau}_{fy}\cdot\overline{\tau}_{gy}+\overline{\tau}_{fx}\cdot\overline{\tau}_{gx}\right)\label{eq:rotation}\end{multline}
where $\arctan\left(y,x\right)\in\left]-\pi,\pi\right]$ is the angle
of a vector $\left[x,y\right]^{T}$ with the $x$-axis. Similarly,
it can be shown that the mirroring that minimizes $\left|\overline{f}-S\left(\overline{g}\right)\right|$
is the composition of a rotation of angle\begin{multline}
\beta=\arctan\\
\left(\overline{\tau}_{fx}\cdot\overline{\tau}_{gy}+\overline{\tau}_{fy}\cdot\overline{\tau}_{gx}\;,\;\overline{\tau}_{fy}\cdot\overline{\tau}_{gy}-\overline{\tau}_{fx}\cdot\overline{\tau}_{gx}\right)\label{eq:mirroring}\end{multline}
 by a swap of the sign of the $x$-coordinates. Equations \ref{eq:rotation}
and \ref{eq:mirroring} allow to compute \ref{eq:rotatron-dist}.

\begin{figure}
\includegraphics[width=8cm]{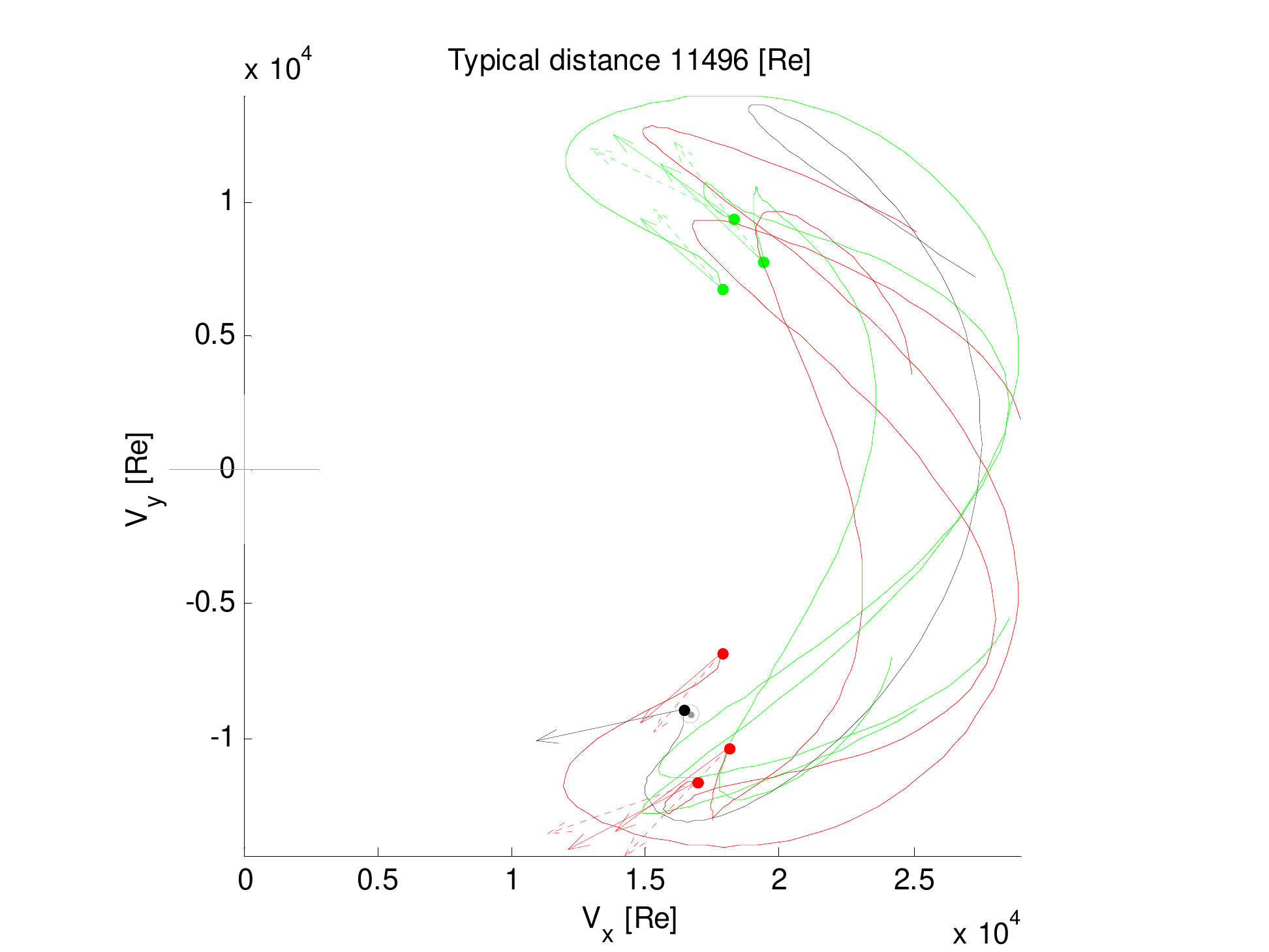}

\caption{A trajectory and its neighbors in terms of rotatron distance. Smooth
curve: Laguerre approximation of trajectory, stippled arrow: true
force, smooth arrow: predicted force. Black is for the trajectory
used to enter the model to find the force. Green and red are used
for the three closest points in the database, respectively before
and after rotation or mirroring.}
\label{Flo:neighbours}
\end{figure}

\noindent Figure \ref{Flo:neighbours} shows a trajectory and the
trajectories within a small database that have the smallest distance
to it, measured using $d$.%

\subsection{\noindent Fractal dimension}

\noindent Considering a relatively uniform cloud of points, the number
$m$ of points in a sphere is proportional to the radius $r$ of the
sphere to the power of $p$, where $p$ is the dimension of the space
in which the cloud is defined. For example, using $2\times10$ coefficients
to describe both components of a trajectory, if the database was filling
this space, the number of points within the sphere would be $m\propto r^{20}$
(no realistic experimental database can {}``fill'' such a volume).

\noindent Conversely, one can define the fractal dimension (or Minkowski-Bouligand
dimension \cite{mandelbrot67}) $p$ of a set (in particular, of a
{}``database'' of Laguerre coefficients) by counting the number
$m(r)$ of pairs of points in the set which have a distance smaller
than $r$:\begin{equation}
p(r)\equiv\frac{\partial\log m(r)}{\partial\log r}\end{equation}
One should either smooth $m(r)$ or compute the derivative by finite
differences over a large enough interval. Note that the fractal dimension
$p$ is a function of the scale $r$.

\noindent Imagine that we have a series of data-points $(x,y,z)$,
and we are investigating whether $z$ can be predicted using $x$
and $y$. Let us imagine that the fractal dimension of the set of
$(x,y)$ pairs is $2$ (the set of $(x,y)$ fills the plane). If the
fractal dimension of the set of $(x,y,z)$ is equal to $2$, then
the set of $(x,y,z)$ is within a surface, and $z$ can be predicted
using $x$ and $y$. If the fractal dimension of the set of $(x,y,z)$
is equal to $3$, the data forms a cloud, and $x$ and $y$ are not
sufficient to predict $z$, other hidden variables must be at play.
These concepts are now applied to the study of the database.

\begin{figure}
\includegraphics[width=9cm]{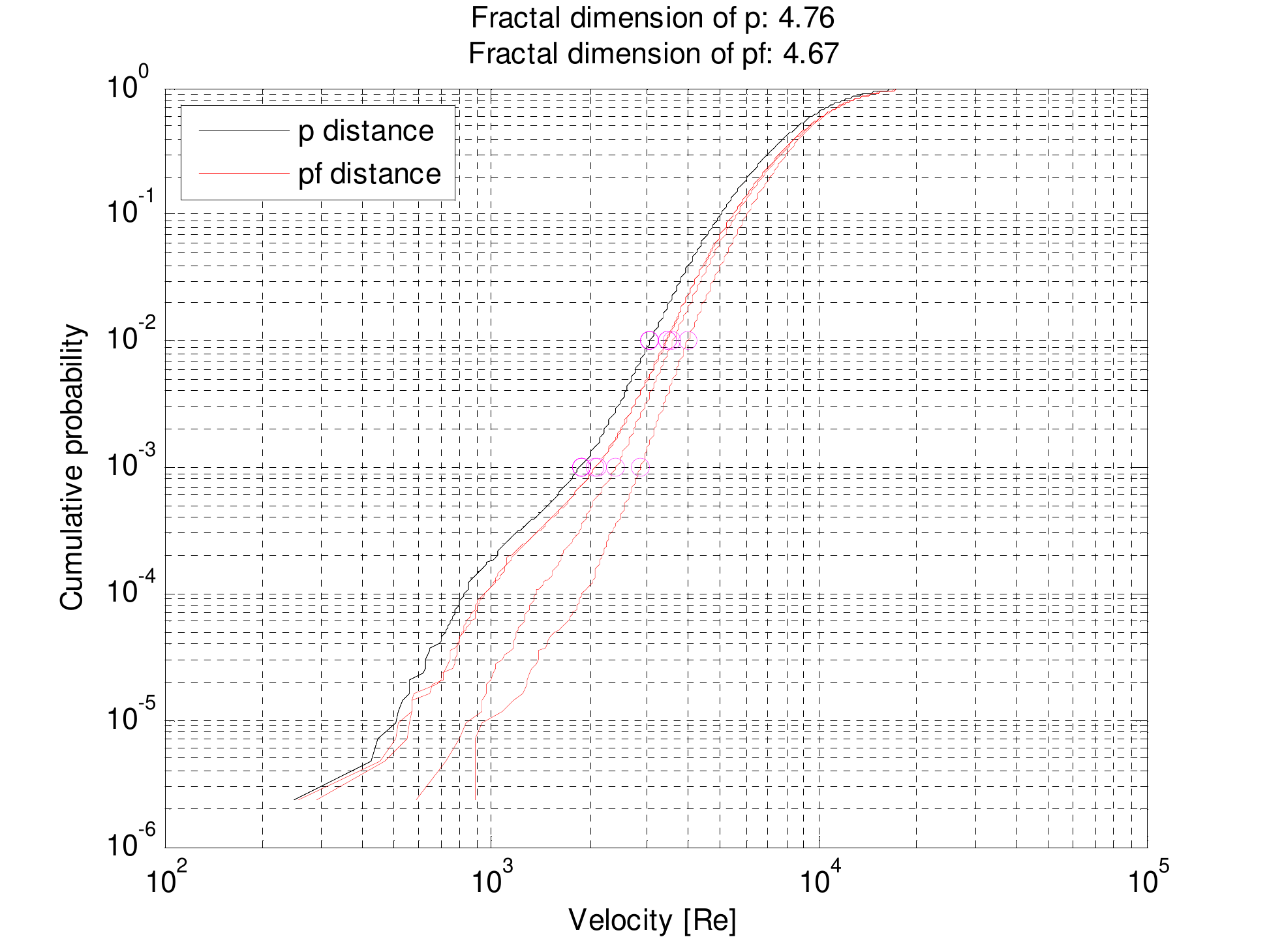}\caption{Computing fractal dimensions}
\label{Flo:fradim}
\end{figure}

\noindent Figure \ref{Flo:fradim} shows the cumulative distribution
of the distances between trajectories (black curve) computed using
Equation \ref{eq:rotatron-dist}. $p$ is seen to depend on the scale
on the scale: from afar ($r>2\times10^{4}\;[ilr\, Re]$), the slope
of the curve is zero, hence the dimension is zero: all the data are
lumped into a point. Zooming into the data set ($r=3\times10^{3}\;[ilr\, Re]$)
one can discern a cloud of dimension $4.76$. At $r=1.5\times10^{3}\;[ilr\, Re]$
the slope decreases to about $p=2$, and it is believed that this
is the dimension of the dataset for a given point along the riser.
At small scale ($r<1\times10^{3}\;[ilr\, Re]$), the dimension increases
again, possibly due to noise in the data. or weaknesses in the Laguerre
approximation.

\noindent The red curves in Figure \ref{Flo:fradim} are computed
by adding the sum of squares of the differences between force components
(suitably scaled) to the squares of the distances between trajectories,
and then extracting the square root. The four red curves are drawn
using the original force data, to which Gaussian noise of standard
deviation $0,$ $10^{7}$, $10^{8}$ and $10^{9}\:[N/m]$ respectively
has been added. The standard deviation of the original force is about
$2\times10^{8}\:[N/m]$ . The two first red curves are indistinguishable,
which seems to indicate that we cannot expect to achieve a 10\% precision
in force predictions. The marked difference with curves 3 and 4 shows
however that we have assets in hand to predict the force. Similar
curves have been produced with added noise of standard deviations
$1\times10^{7}$, $2\times10^{7}$... $10\times10^{7}\;[N/m]$, and
already at \textrm{$2\times10^{7}\;[N/m]$} the curve is distinct
from the one based on the original data. 

\noindent %

\section{Dynamic analysis\label{sec:Dynamic-analysis}}

\subsection{Foreword}

Once it is possible to predict hydrodynamic forces on a cross section
for a given velocity history, the next development is to include the
force thus predicted in a dynamic time domain simulation. Because
the VIV forces introduce severe non-linearities, a naive connection
(where the forces are just added to the right hand side of the system)
might lead to slow convergence, or to divergence of the Newton-Raphson
iterations used at each time step. To obtain a proper formulation,
it is necessary to jointly treat the system of differential equations
composed of the state equations of the structure, and the differential
equations (\ref{eq:dual-diffeq}) followed by the Laguerre coefficient.
However in doing so, for each displacement degree of freedom, $n$
Laguerre coefficients are added, and it is crucial for efficiency
to eliminate them before solving a large linear system of equations.

To this effect, in this Section, the following sequence of transformations
is applied to the differential equations: 
\begin{enumerate}
\item The differential equations are first set in incremental form (Section
\ref{sub:Incremental-form}).
\item Time discretization by the Newmark-$\beta$ method is introduced (Section
\ref{sub:Time-discretisation}). 
\item The Laguerre coefficients are condensed out of the system of equations
(Section \ref{sub:Condensation}). 
\item Finite element interpolation is introduced (space discretisation),
using Gauss quadrature (Section \ref{sub:Spacial-discretisation}).
\end{enumerate}
This particular sequence leads to a VIV model that is implemented
at the Gauss point level, and can easily be introduced in a general
purpose FEM software with standard, displacement based beam or cable
elements. Another sequence, 1, 4, 2, 3, can be used to obtain either
a hybrid element, or alternatively, a mixed element which would require
a specialized solver for optimal efficiency. These alternatives are
more difficult to integrate into existing software working with displacement
based elements, and are not discussed here.

\subsection{Differential equations\label{sub:Differential-equations}}

The dynamic differential equation of a 3D beam subjected to VIV loads
can be formalized as\begin{multline}
r_{di}\left(x_{[bj]},\dot{x}_{[bj]},\ddot{x}_{[bj]},t\right)\\
=\lambda_{Nm^{-1}}^{-1}\hat{f}_{d}\left(\tau_{[pb]i}\right)+E_{di}\label{eq:state}\end{multline}
where Newton's {}``dot'' notation for a time derivative stands for
a derivation with respect to \emph{unscaled} time $t$, as opposed
to scaled time $t^{*}$, and with \cite{morison50,faltinsen90}\begin{eqnarray}
E_{di} & = & C_{L}\,\rho\nu\left(\dot{w}_{di}-\dot{x}_{di}\right)\nonumber \\
 & + & C_{Q}\,\frac{1}{2}\rho D_{i}\left|\dot{w}_{di}-\dot{x}_{di}\right|\left(\dot{w}_{di}-\dot{x}_{di}\right)\nonumber \\
 & + & C_{M}\frac{\pi}{4}\rho D_{i}^{2}\left(\ddot{w}_{di}-\ddot{x}_{di}\right)\nonumber \\
 & + & \hphantom{C_{M}}\frac{\pi}{4}\rho D_{i}^{2}\ddot{w}_{di}\end{eqnarray}
The four terms in the above Morison's equation are the linear drag,
the quadratic drag, the sum of diffraction and added mass forces,
and the Froude-Krylov forces. The fourth term introduces the correction
discussed in Section \ref{sub:Frode-Krylov-forces}. 

If $C_{L}$, $C_{Q}$ or $C_{M}$ are set to values different from
zero, then it is necessary to substract the correspond values from
the forces $f_{di}$ used to train the rotatron. Experience shows
that the Using $C_{M}=1$, $C_{Q}=1$ and $C_{L}=0$ contributes to
the stability of the dynamic analysis.

Equation \ref{eq:dual-diffeq} must be scaled to keep only derivatives
with respect to unscaled time, for the application of Newmark-$\beta$
(Section \ref{sub:Time-discretisation}) \begin{equation}
\frac{\partial\tau_{lbi}}{\partial t^{*}}=\mu_{lp}\tau_{pbi}+n_{l}\lambda_{ms}^{-1}\left(\dot{x}_{bi}-\dot{w}_{bi}\right)\label{eq:lag-1}\end{equation}
so that\begin{equation}
\lambda_{s}^{-1}\dot{\tau}_{lbi}=\mu_{lp}\tau_{pbi}+n_{l}\lambda_{ms^{-1}}\left(\dot{x}_{bi}-\dot{w}_{bi}\right)\label{eq:lag-2}\end{equation}

The indices $d$ and $b$ span pairs of directions, orthogonal to
the cylinder. Indices $i$ and $j$ stand for positions along the
cylinder, and span a continuous set of values (coordinates along the
cylinder). Indices $l$ and $p$ refer to the Laguerre coefficients
of various degrees. Forces $\hat{f}_{di}=\hat{f}_{d}\left(\tau_{[pb]i}\right)$
at location $i$ only depend on the Laguerre coefficients $\tau_{pbi}$
for the same location. At that location, the force component in direction
$d$ depend on the Laguerre coefficients of all degrees $b$ for both
directions $p$. $\rho$ is the fluid density $\ddot{w}_{di}$ is
the acceleration of the undisturbed fluid. $\frac{\pi}{4}\rho D_{i}^{2}\ddot{w}_{di}$
stands for the Froude-Krylov forces. Diffraction forces are present
in the laboratory tests and hence accounted for by $\hat{f}_{d}$.

\subsection{Incremental form\label{sub:Incremental-form}}

The incremental form of Equations \ref{eq:state} and \ref{eq:lag-2}is
\begin{multline}
r_{di}+k_{dibj}dx_{bj}+c_{dibj}d\dot{x}_{bj}+m_{dibj}d\ddot{x}_{bj}\\
=\lambda_{Nm^{-1}}^{-1}\hat{f}_{di}+h_{dipbj}d\tau_{pbj}+E_{di}\label{eq:incstate}\end{multline}
\begin{multline}
\lambda_{s}^{-1}\left(\dot{\tau}_{lbi}+d\dot{\tau}{}_{lbi}\right)=\mu_{lp}\left(\tau_{pbi}+d\tau_{pbi}\right)\\
+n_{l}\lambda_{ms^{-1}}\left(\dot{x}_{bi}+d\dot{x}{}_{bi}-\dot{w}_{di}\right)\label{eq:inclag-2}\end{multline}
with\begin{eqnarray}
k_{dibj} & = & \dfrac{\partial r_{di}}{\partial x_{bj}}\\
c_{dibj} & = & \dfrac{\partial r_{di}}{\partial\dot{x}_{bj}}+C_{Q}\rho D_{i}\delta_{ij}\left[\vphantom{\frac{A}{A}}\left|\dot{w}_{[p]i}-\dot{x}_{[p]i}\right|\delta_{bd}\right.\nonumber \\
 &  & +\left.\vphantom{\frac{A}{A}}\left(\dot{w}_{di}-\dot{x}_{di}\right)\left(\dot{w}_{bi}-\dot{x}_{bi}\right)\left|\dot{w}_{[p]i}-\dot{x}_{[p]i}\right|^{-1}\right]\nonumber \\
 &  & +C_{L}\rho\nu\delta_{ij}\delta_{bd}\\
m_{dibj} & = & \dfrac{\partial r_{di}}{\partial\ddot{x}_{bj}}+C_{M}\frac{\pi}{4}\rho D_{i}\delta_{ij}\delta_{bd}\\
h_{dipbj} & = & \lambda_{Nm^{-1}}^{-1}\dfrac{\partial\hat{f}_{di}}{\partial\tau_{pb}}\delta_{ij}\end{eqnarray}
The expression for $\frac{\partial\hat{f}_{di}}{\partial\tau_{pb}}$
is presented in Appendix \ref{sec:Rotatron-gradients}.

\subsection{Time discretization\label{sub:Time-discretisation}}

Newmark-$\beta$ is a method geared towards 2nd order differential
equations. Equation \ref{eq:inclag-2}, however, is only of the first
order, and this opens two options: we can treat Equation \ref{eq:inclag-2}
as being of the second order in $\tau_{lbi}$, but with the coefficient
of $\ddot{\tau}_{lbi}$ being zero. Alternatively, we can introduce
the antiderivative $T_{lbi}$ of $\tau_{lbi}$, and treat Equation
\ref{eq:inclag-2} as being of the second order in $T_{lbi}$, but
with the coefficient of $T_{lbi}$ being zero. The later option was
chosen, based on the weak justification that this treats $\tau_{lbi}$
and $\dot{x}_{bj}$ both as first derivatives, which seems natural
considering Equation \ref{eq:tadpole}.

Applying Newmark-$\beta$ to Equations \ref{eq:incstate} and \ref{eq:inclag-2}
in this way yields\begin{multline}
\forall d,i,\;\left[k_{dibj}+\frac{\gamma}{\beta dt}c_{dibj}+\frac{1}{\beta dt^{2}}m_{dibj}\right]dx_{bj}\\
-\frac{\gamma}{\beta dt}h_{dipbj}dT_{pbj}\\
=\lambda_{Nm^{-1}}^{-1}\hat{f}_{di}+E_{di}-r_{di}\\
+c_{dibj}b_{bj}^{x}+m_{dibj}a_{bj}^{x}-h_{dipbj}b_{pbj}^{\tau}\label{eq:incstate-1}\end{multline}
and\begin{multline}
-\frac{\gamma}{\beta dt}n_{l}\lambda_{ms^{-1}}dx_{bi}+\left[\frac{1}{\beta dt^{2}}\lambda_{s}^{-1}\delta_{lp}-\frac{\gamma}{\beta dt}\mu_{lp}\right]dT_{pbi}\\
=n_{l}\lambda_{ms^{-1}}\left(\dot{x}_{bi}-\dot{w}_{bi}\right)+\mu_{lp}\tau_{pbi}-\lambda_{s}^{-1}\dot{\tau}_{lbi}\\
-n_{l}\lambda_{ms^{-1}}b_{bi}^{x}-\mu_{lp}b_{pbi}^{\tau}+a_{lbi}^{\tau}\label{eq:inclag-1}\end{multline}
with\begin{eqnarray}
a_{bj}^{x} & = & \frac{1}{\beta dt}\dot{x}_{bj}+\frac{1}{2\beta}\ddot{x}_{bj}\\
b_{bj}^{x} & = & \frac{\gamma}{\beta}\dot{x}_{bj}+\left(\frac{\gamma}{2\beta}-1\right)dt\,\ddot{x}_{bj}\\
a_{pbj}^{\tau} & = & \frac{1}{\beta dt}\tau_{pbj}+\frac{1}{2\beta}\dot{\tau}_{pbj}\\
b_{pbj}^{\tau} & = & \frac{\gamma}{\beta}\tau_{pbj}+\left(\frac{\gamma}{2\beta}-1\right)dt\,\dot{\tau}_{pbj}\end{eqnarray}
For refinement iterations, $a_{bj}^{x}$, $b_{bj}^{x}$, $a_{pbj}^{\tau}$
and $b_{pbj}^{\tau}$ are set to zero. Typicaly, $\gamma=$$\frac{1}{2}$,
$\beta=\frac{1}{4}$. The step $dt$ refers to unscaled time.

As usual in the Newmark-$\beta$ method, the increments for the time
derivatives are found from the increment as\begin{eqnarray}
d\dot{x}_{bj} & = & \frac{\gamma}{\beta dt}dx_{bj}-b_{bj}^{x}\\
d\ddot{x}_{bj} & = & \frac{1}{\beta dt^{2}}dx_{bj}-a_{bj}^{x}\\
d\tau_{pbj} & = & \frac{\gamma}{\beta dt}dT_{pbj}-b_{pbj}^{\tau}\label{eq:inc tau dot}\\
d\dot{\tau}_{pbj} & = & \frac{1}{\beta dt^{2}}dT_{pbj}-a_{pbj}^{\tau}\label{eq:inc tau ddot}\end{eqnarray}

\subsection{Condensation\label{sub:Condensation}}

The time discrete equations can be rewritten in a compact form:\begin{eqnarray}
s_{dibj}^{1}dx_{bj}-s_{dipbj}^{2}dT_{pbj} & = & s_{di}^{3}\\
s_{l}^{4}dx_{bi}+s_{lp}^{5}dT_{pbi} & = & s_{lbi}^{6}\end{eqnarray}
with\begin{eqnarray}
s_{dibj}^{1} & = & k_{dibj}+\frac{\gamma}{\beta dt}c_{dibj}+\frac{1}{\beta dt^{2}}m_{dibj}\\
s_{dipbj}^{2} & = & \frac{\gamma}{\beta dt}h_{dipbj}\\
s_{di}^{3} & = & \lambda_{Nm^{-1}}^{-1}\hat{f}_{di}+E_{di}-r_{di}\nonumber \\
 &  & +c_{dibj}b_{bj}^{x}+m_{dibj}a_{bj}^{x}\nonumber \\
 &  & -h_{dipbj}b_{pbj}^{\tau}\\
s_{l}^{4} & = & -\frac{\gamma}{\beta dt}n_{l}\lambda_{ms^{-1}}\\
s_{lp}^{5} & = & \frac{1}{\beta dt^{2}}\lambda_{s}^{-1}\delta_{lp}-\frac{\gamma}{\beta dt}\mu_{lp}\label{eq:s5}\\
s_{lbi}^{6} & = & n_{l}\lambda_{ms^{-1}}\left(\dot{x}_{bi}-\dot{w}_{bi}\right)+\mu_{lp}\tau_{pbi}-\lambda_{s}^{-1}\dot{\tau}_{lbi}\nonumber \\
 &  & -n_{l}\lambda_{ms^{-1}}b_{bi}^{x}-\mu_{lp}b_{pbi}^{\tau}+\lambda_{s}^{-1}a_{lbi}^{\tau}\end{eqnarray}

One can then condense $dT_{pbi}$ out of the above system of equations:\begin{equation}
dT_{pbj}=\left(s^{5}\right)_{pl}^{-1}\left(s_{lbj}^{6}-s_{l}^{4}dx_{bj}\right)\label{eq:laguerre-increment}\end{equation}
\begin{multline}
\left[s_{dibj}^{1}+s_{dipbj}^{2}\left(s^{5}\right)_{pl}^{-1}s_{l}^{4}\right]dx_{bj}\\
=s_{di}^{3}+s_{dipbj}^{2}\left(s^{5}\right)_{pl}^{-1}s_{lbj}^{6}\label{eq:condensed material}\end{multline}
Equation \ref{eq:condensed material} is forced into {}``Newmark''
form as\begin{multline}
\left[k_{dibj}^{*}+k_{dibj}+\frac{\gamma}{\beta dt}c_{dibj}+\frac{1}{\beta dt^{2}}m_{dibj}\right]dx_{bj}\\
=\hat{f}_{di}^{*}-r_{di}+c_{dibj}b_{bj}^{x}+m_{dibj}a_{bj}^{x}\label{eq:newmark material}\end{multline}
with\begin{align}
k_{dibj}^{*} & =s_{dipbj}^{2}\left(s^{5}\right)_{pl}^{-1}s_{l}^{4}\\
\hat{f}_{di}^{*} & =\lambda_{Nm^{-1}}^{-1}\hat{f}_{di}+E_{di}-h_{dipbj}b_{pbj}^{\tau}\nonumber \\
 & +s_{dipbj}^{2}\left(s^{5}\right)_{pl}^{-1}s_{lbj}^{6}\end{align}
$k_{dibj}^{*}$and $\hat{f}_{di}^{*}$ both depend on $dt$, $\beta$
and $\gamma$: The symbol $k_{dibj}^{*}$ was chosen to indicate that
the matrix is handled by the Newmark-$\beta$ solver in the same way
as a stiffness, however this terms can not be interpreted physically
as a stiffness.

\subsection{Spacial discretization\label{sub:Spacial-discretisation}}

The consistent discretisation by Galerkin finite elements of Equation
\ref{eq:newmark material} leads to \begin{eqnarray}
K_{nm}^{*} & = & N_{din}k_{dibj}^{*}N_{bjm}\\
\hat{F}_{n}^{*} & = & N_{din}\hat{f}_{di}^{*}\end{eqnarray}
$K_{nm}^{*}$ and $\hat{F}_{n}^{*}$ are typicaly computed by Gauss
quadrature. Note that no space derivative is present in $k_{dibj}^{*}$,
so no partial integration or Gauss quadrature with curvature shape
function appears. One can hence simplify the expression of the element
matrix to \begin{equation}
K_{nm}^{*}=N_{din}k_{dib}^{*}N_{bim}^{}\end{equation}
which means {}``same quadrature as for a mass matrix''.

\subsection{Implementation}

In non-linear FEM code, incremental matrices and vectors are computed
by Gauss quadrature. The Gauss quadrature involves shape functions,
tensors that are local, continuous versions of the stiffness, damping
and mass matrices, and the force imbalance vector. For example for
the drag damping of a beam element, the tensor relates a vector which
components are increments in velocities in three directions, to another
vector which components are increments in forces per unit length in
three directions. 

Within an iteration, the linear solver provides incremental nodal
positions, velocities and accelerations for the model. These are disassembled
and provided to the elements. The elements compute positions, velocities
and accelerations (and more) in a co-rotated reference system at Gauss
points. The resulting values are handed to the VIV-Gauss point procedure.

The axial velocities are discarded. The procedure scales the provided
values using Equations \ref{eq:scale_velocity} and \ref{eq:scale_acceleration}.

Having stored the previous approximation of the scaled position, the
procedure determines the position increment $dx_{bj}$, and then uses
Equation \ref{eq:laguerre-increment} to obtain the Laguerre coefficient
increment $dT_{pbj}$. From there, Equations \ref{eq:inc tau dot}
and \ref{eq:inc tau ddot} are used to compute $d\tau_{pbj}$ and
$d\dot{\tau}_{pbj}$ . The values of $T_{pbj}$, $\tau_{pbj}$ and
$\dot{\tau}_{pbj}$ are updated from previously stored values. $\tau_{pbj}$
is then used to evaluate $\hat{f}_{di}$ and its derivative with respect
to $\tau_{pbj}$. These are scaled back, and Froude-Krylov forces
are added, leading to $k_{dibj}^{*}$ and \textrm{$\hat{f}_{di}^{*}$. }

The above matrix and vector are padded with zeros to indicate zero
force in the axial direction and zero torque.

The condensation of a larger system of time-discretized equations
introduces some inelegant features compared to standard dynamic FEM:
the VIV-Gauss point must be provided with $\beta$, $\gamma$ and
$dt$ and a flag showing wether a call is made at a step or within
a refinement iteration. 

Note that the dynamic FEM computation does not make any use of the
recursive filter presented in Section \ref{sub:Recursive-filter}:
this filter is used only in the training of the rotatron model. In
the context of training, the filter had the advantage of making refinement
iterations unnnecessary. It further allowed to avoid using Newmark-$\beta$
in a situation with prescribed displacement, for which it is not well
suited.

\section{\noindent Results\label{sec:Application}}

\subsection{Training\label{sub:Training-1}}

\begin{figure}
\includegraphics[width=9cm]{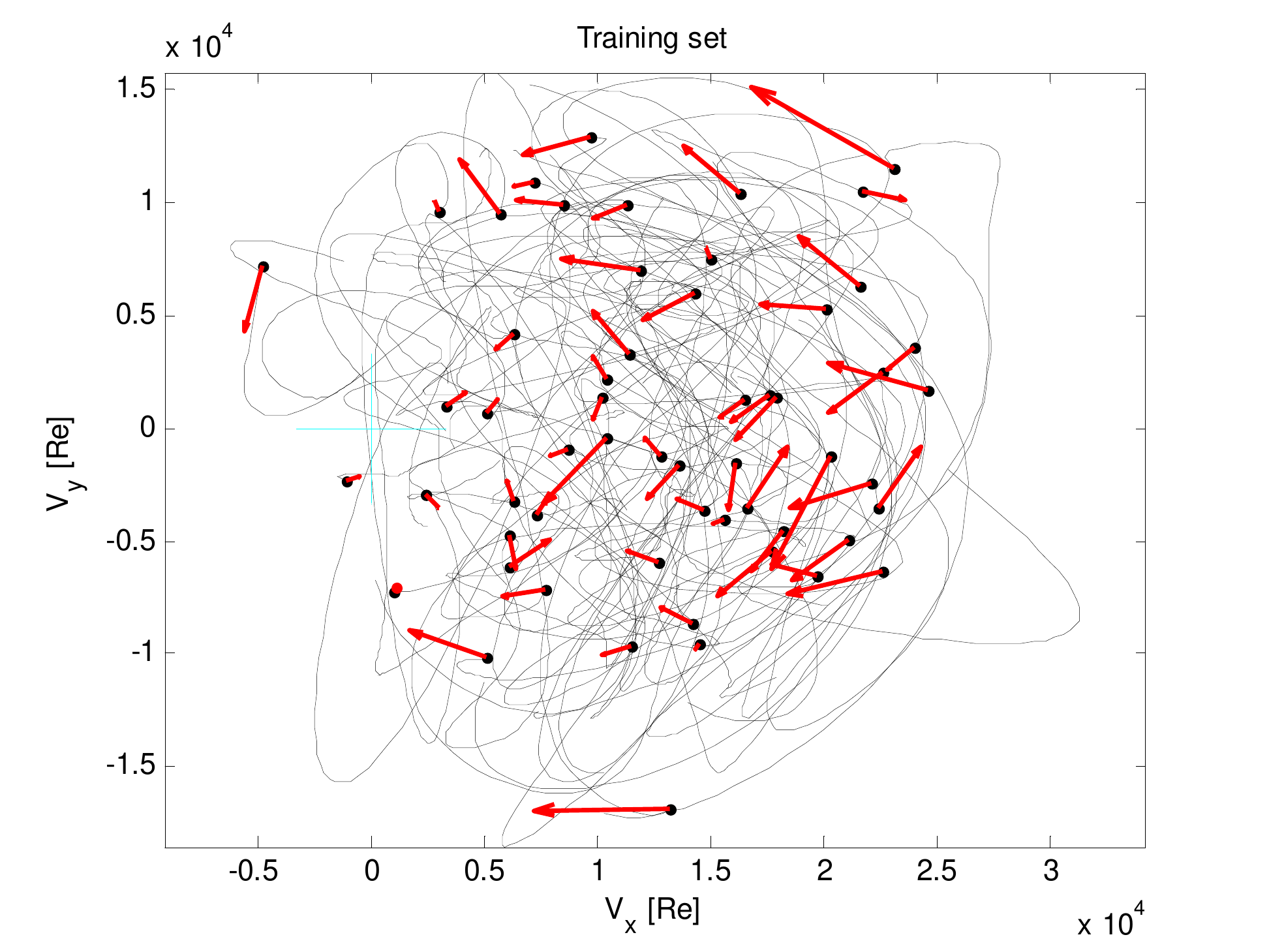}\caption{Force vector and Laguerre approximation of velocity, for a fraction
(1\%) of the data used to train the rotatron. The blue cross marks
the origin (zero velocity relative to water)}
\label{Flo:decao_database}
\end{figure}

The Norwegian Deepwater Program was a research effort in which reduced
scale tests were carried out on long, flexible riser models, subject
to uniform or sheared current \cite{braaten05}. The displacement
histories thus aquired at 19 points along the riser model were later
prescribed on short stiff cylinders, and the hydrodynamic forces acting
on the cylinders directly measured \cite{decao10}. The data from
\cite{decao10} that is used in this work consists of the displacements
at 19 points along the NDP riser model, for 3 current profiles (Table
\ref{Flo:NDP-Re}), so a total of 57 short cylinder runs. For each
of the 57 runs, 100 instants are randomly selected, yielding a training
set of the rotatron with 5700 {}``points''. Each {}``point'' consistes
of two sets of $n=30$ Laguerre coefficients and the two components
of the corresponding force (Figure \ref{Flo:decao_database}). 

\begin{table}
\noindent \centering{}\begin{tabular}{lrr}
\hline 
Test name & Reynolds number & Current\tabularnewline
\hline 
TN2030 & 13500 & uniform\tabularnewline
TN2340 & 0-16200 & shear\tabularnewline
TN2370 & 0-24300 & shear\tabularnewline
\hline
\end{tabular}\caption{Reynolds number in NDP tests}
\label{Flo:NDP-Re}
\end{table}

The rotatron was trained using $n=30$ Laguerre polynomials, 200 neurons
in the hidden layer, and 50 to 1000 iterations of the conjugate gradient
optimization algorithm. 

\begin{figure*}
\includegraphics[bb=500bp 100bp 1500bp 1000bp,clip,height=24cm]{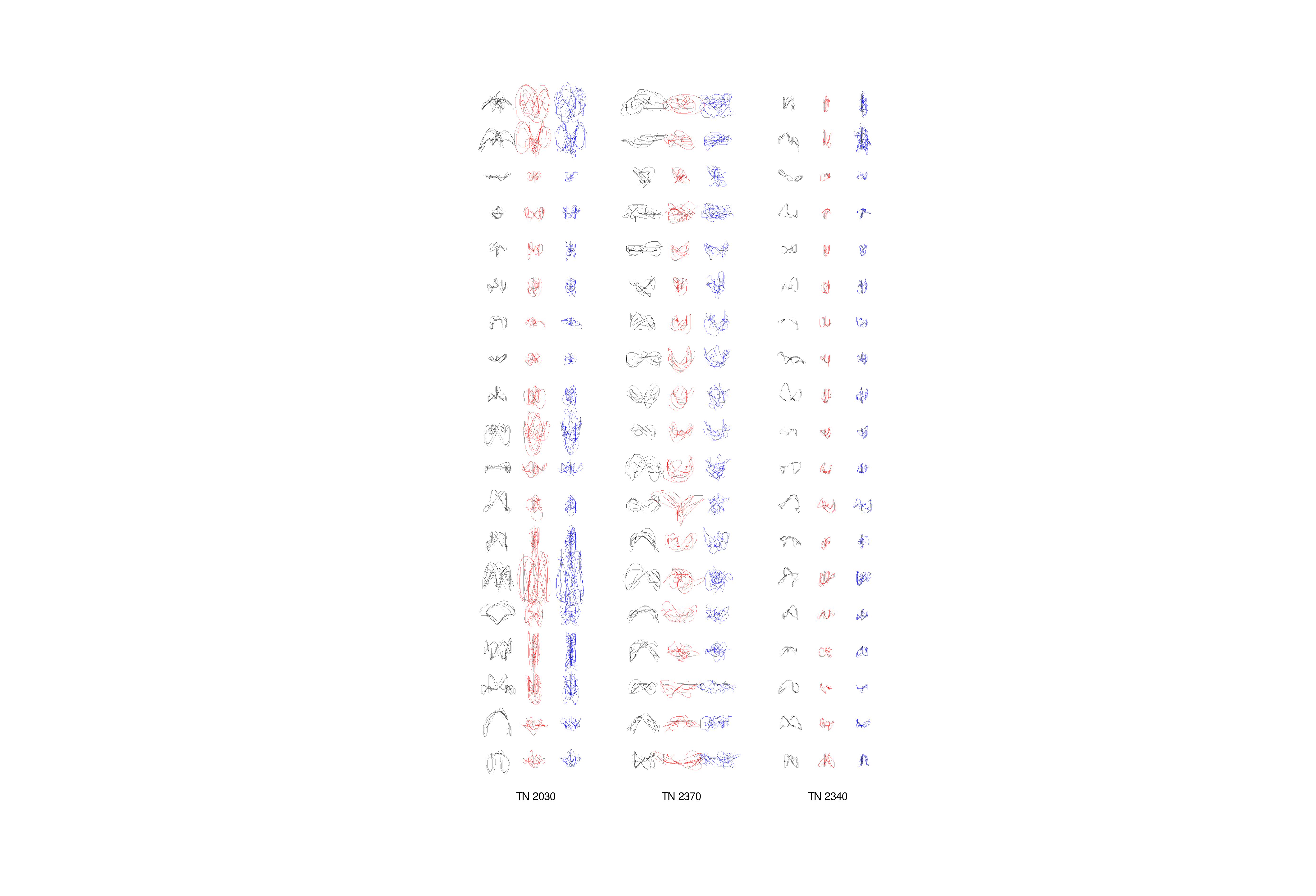}\caption{Quality of prediction on the training data set. Velocity (black),
training force (red), predicted force (blue). All velocities and all
forces presented at the same scales}
\label{Flo:decao_reproduction}
\end{figure*}

Figure \ref{Flo:decao_reproduction} shows how the rotatron predicts
the forces for the trajectories in the above-mentionned 57 runs of
short cylinder tests. the comparison of the forces aquired experimentaly
with the forces predicted using the model. The model's ability to
predict these forces seems to be good, although we lack a good criteria
to judge that yet.%

\begin{figure}
\includegraphics[bb=60bp 0bp 450bp 420bp,clip,width=8cm]{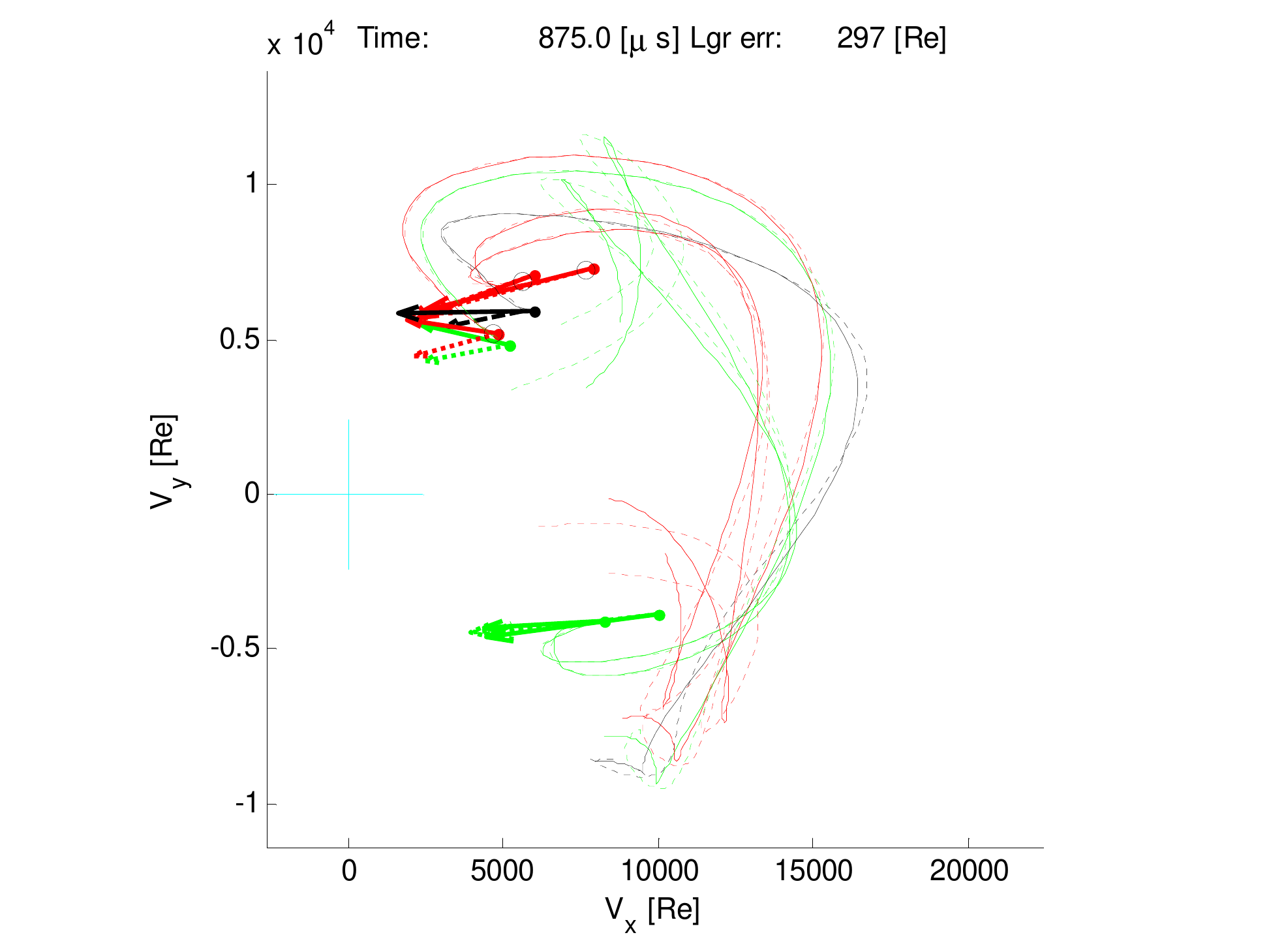}\caption{Illustration of the approximation process. See Section \vref{sub:Training-1}}
\label{Flo:total_view}
\end{figure}

Figure \ref{Flo:total_view} provides a visualisation of the different
steps of the modelisation process, and is hence a useful diagnostic
tool.  It shows:
\begin{description}
\item [{Stippled\textcolor{white}{\_}black\textcolor{white}{\_}line}] The
trajectory for which a force prediction is wanted.
\item [{Smooth\textcolor{white}{\_}black\textcolor{white}{\_}line}] The
Laguerre approximation to the above trajectory, used to enter the
rotatron.
\item [{Stippled\textcolor{white}{\_}black\textcolor{white}{\_}arrow}] The
measured force for the above trajectory.
\item [{Smooth\textcolor{white}{\_}black\textcolor{white}{\_}arrow}] The
predicted force for the above trajectory.
\item [{Green}] Neighboring (in the sense of the rotatron distance, Equation
\vref{eq:rotatron-dist}) trajectories from experiments, used in the
training set (and corresponding Laguerre approximation, experimentally
measured force and predicted force).
\item [{Red}] Same as the above after rotation and/or mirroring.
\end{description}
Figure \ref{Flo:total_view} gives an indication of the quality of
the Laguerre approximation, the adequacy of the training set for the
trajectory at hand, the presence of contradictions in the training
set near the trajectory at hand, the quality of the fit of the rotatron
to the training data and finally the quality of the interpolation
between training points.

\subsection{Dynamic analysis of a flexible riser}

VIV depends not only of current velocities, but on the type of slender
system they act upon. Tension, stiffness, damping, length and boundary
conditions affect the vibration and hence the velocity trajectories
that appear in the vibration. Hence the database used in Section \ref{sub:Training-1}
to train the rotatron is specialised, not only to a few Reynolds numbers
for the current velocity, but also to some extend to the particular
model used in the NDP program. No study was carried out in this work
on which changes in the structure change its way of vibrating to the
point were the rotatron provides poor for estimations for it.

Hence, in order to test the performance of the present VIV model within
a dynamic analysis, the simplest case was considered: the riser model
used in the NDP testing program (\cite{braaten05}, characteristics
in Table \ref{Flo:NDPriser}) was modelled. The present method was
used to analyse the test condition TN2370 ($0-24300\, Re$). Numerical
results were compared with those obtained experimentaly on the NDP
model. A laptop, using one core of a dual core processor, took 20
to 50 seconds to compute one second of riser response.

\begin{table}
\noindent \centering{}\begin{tabular}{lrl}
\hline 
Quantity & Value & \tabularnewline
\hline
length  & $38$ & $m$\tabularnewline
outer diameter & $0.027$ & $m$\tabularnewline
$EI$ & $37.2$ & $Nm^{2}$\tabularnewline
$EA$ & $5.09\cdot10^{5}$ & $N$\tabularnewline
mass & $0.933$ & $kg/m$\tabularnewline
tension & $3000$ & $N$\tabularnewline
\hline
\end{tabular}\caption{Characteristics of the NDP reduced scale riser model}
\label{Flo:NDPriser}
\end{table}

\begin{figure*}
\noindent \centering{}\includegraphics[bb=100bp 270bp 500bp 570bp,clip,width=13cm]{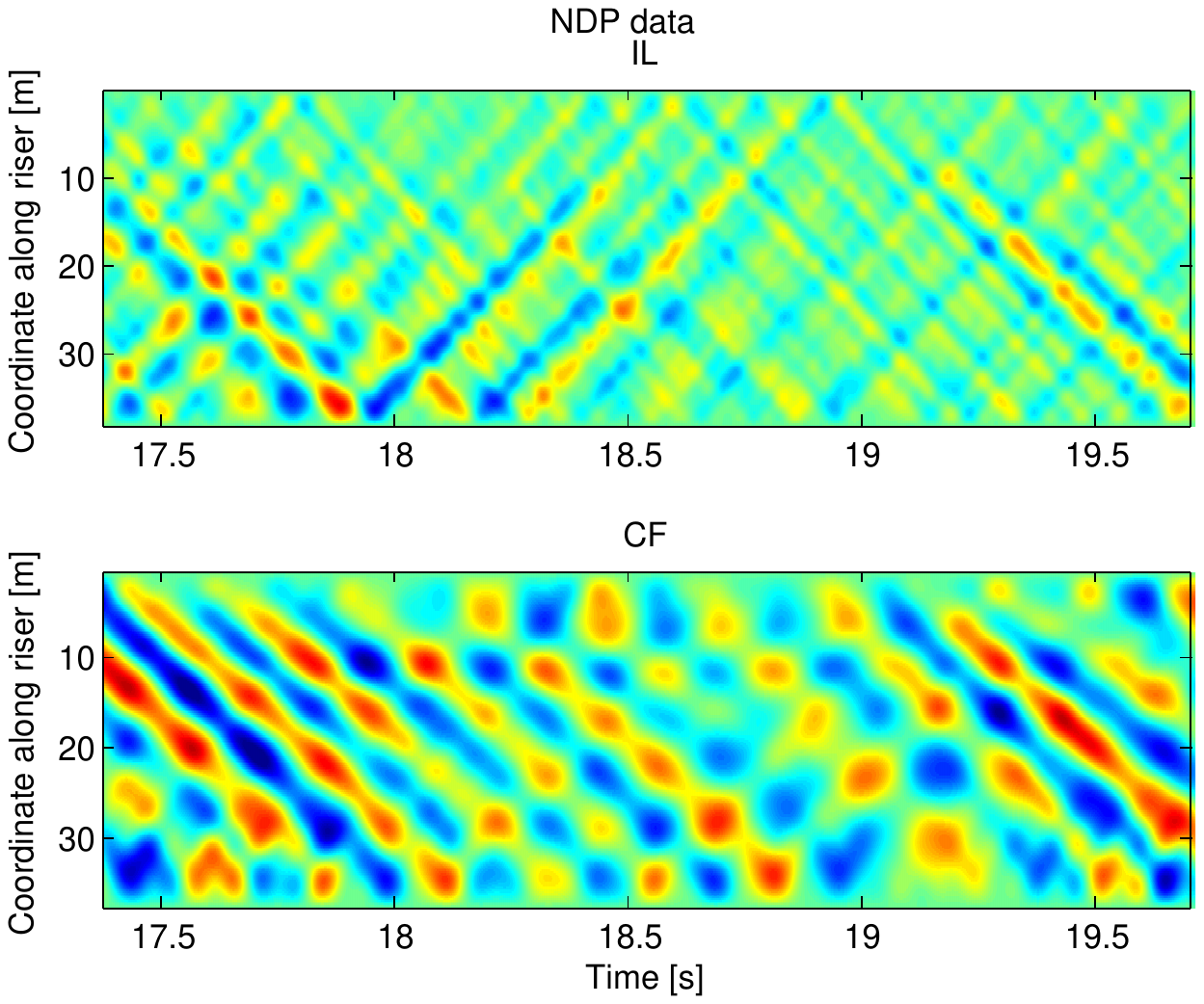}\caption{NDP test results, test 2370 ($0-24300\: Re$). The color coding describes
the displacement}
\label{Flo:NDP2370banner}
\end{figure*}

\begin{figure*}
\noindent \centering{}\includegraphics[bb=100bp 270bp 500bp 570bp,clip,width=13cm]{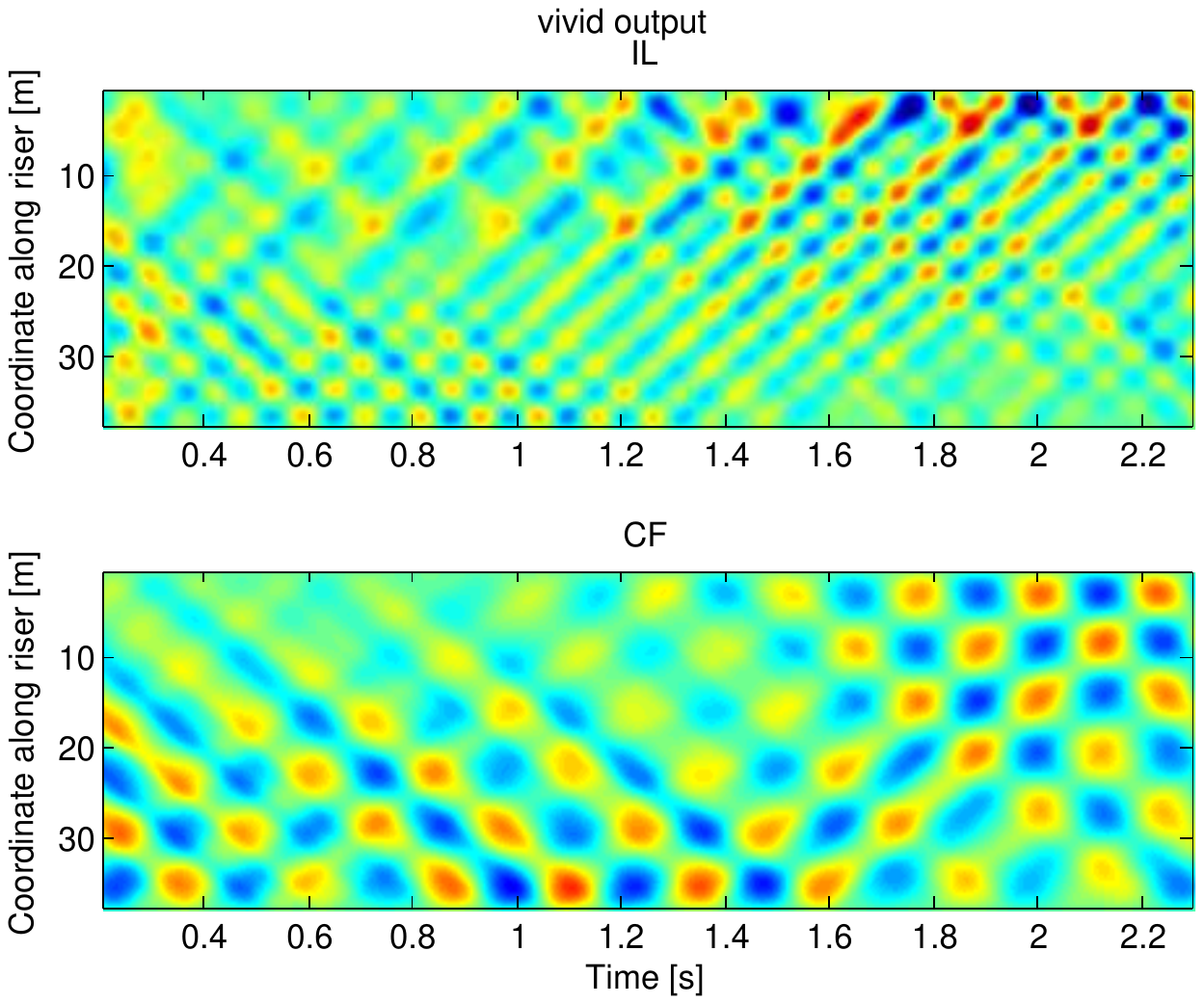}\caption{Analysis results, test 2370 ($0-24300\: Re$)}
\label{Flo:VIVID2370banner}
\end{figure*}

\begin{figure*}
\noindent \centering{}\includegraphics[bb=100bp 270bp 500bp 570bp,clip,width=13cm]{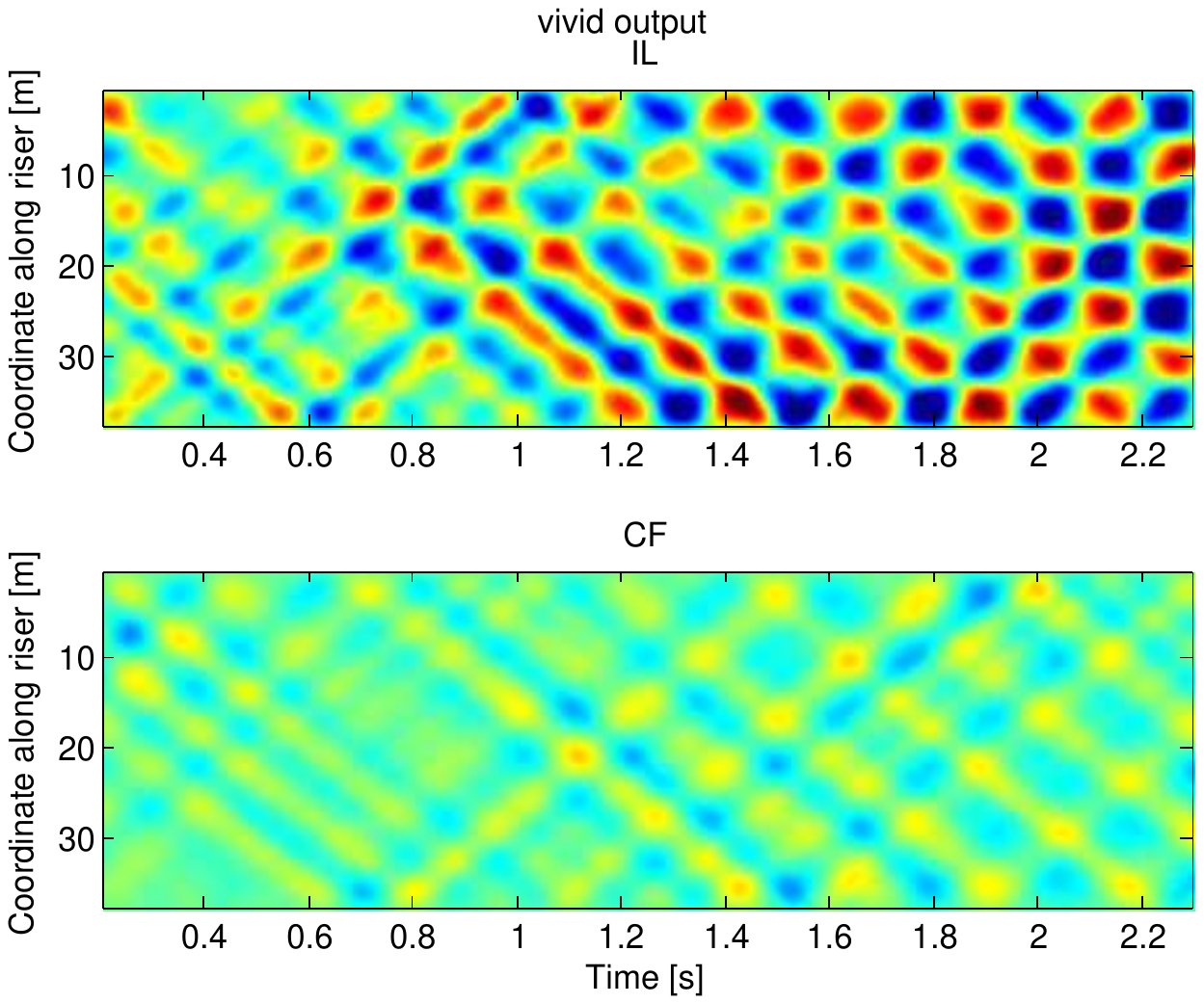}\caption{Analysis results, test 2030 ($13500\, Re$)}
\label{Flo:VIVD2030banner}
\end{figure*}

In Figures \ref{Flo:NDP2370banner} and \ref{Flo:VIVID2370banner},
the horizontal axis is NDP-laboratory time, the vertical axis is the
unscaled length along the riser. The upper subplot shows the response
in line (IL) with the flow, the lower subplot the cross-flow (CF)
response. The color codes the displacements, with the same color scale
used in Figures \ref{Flo:NDP2370banner} to \ref{Flo:VIVD2030banner}. 

Figures \ref{Flo:NDP2370banner} and \ref{Flo:VIVID2370banner} are
for test TN2370 ($0-24300\: Re$). The dynamic simulations captures
the frequency doubling between CF and inline, as well as the instationary
nature of the vibrations. Frequency and amplitude are adequately captured.
The 6th mode's dominance of the CF vibration is correctly captured.
The dynamic analysis assumes constant tension. By contrast, some small
tension modulations seem to displace the position of the lower vibration
node in the test. In test 2370 there is a marked tendency for CF vibrations
to propagate downwards. In the analysis results, CF waves are of a
more static character. The IL vibration in the analysis occurs at
a higher mode (11) than in the test (9). This is best seen by counting
red dots along diagonals, for example starting from coordinate $38m$
and time $17.9s$ in Figure \ref{Flo:NDP2370banner}. As a consequence,
and since, for a given propagation celerity, wavelength and period
are related, the phase drift between IL and CF are of opposite sign
in the analysis and the test.

It proved impossible to reproduce tests 2340 and 2030 in the same
manner. The analysis quickly ends with IL and CF vibrations occurring
at the same frequency, and in Figure \ref{Flo:VIVD2030banner}, with
in-line motions dominating. 

\section{Discussion\label{sec:Discussion}}

\subsection{Force prediction}

Figure \ref{Flo:decao_reproduction} shows that given the velocities
in the training data as input, the model allows to reproduce the forces
in the training data, based on the velocity in the training data. 

The same exercise is carried out with trajectories from another test,
N2430 (not to be confused with TN2340), which was carried out in shear
current ($0\: Re$ to $40500\, Re$ ). In comparison, the highest
current velocity appearing in the training set is $24300\: Re$. In
N2430, the forces are fairly well predicted at the lower current velocities,
while the predictions are very poor at higher velocities. This illustrates
that the present model provides no mechanism to {}``extrapolate''
over Reynolds numbers, in contrast to other VIV models. One could
imagine a future model in which some dependency on the Reynolds number
is encoded, just as symmetries are now encoded in the rotatron. However,
this will not be simple: if for example, viscous forces are to increase
quadraticaly with the Reynolds number and the inertial forces linearly,
how in the first place to partition a hydrodynamic force from a test
in two such components? Would such a scaling work at all?

\begin{figure}
\includegraphics[bb=350bp 70bp 550bp 750bp,clip,width=7cm]{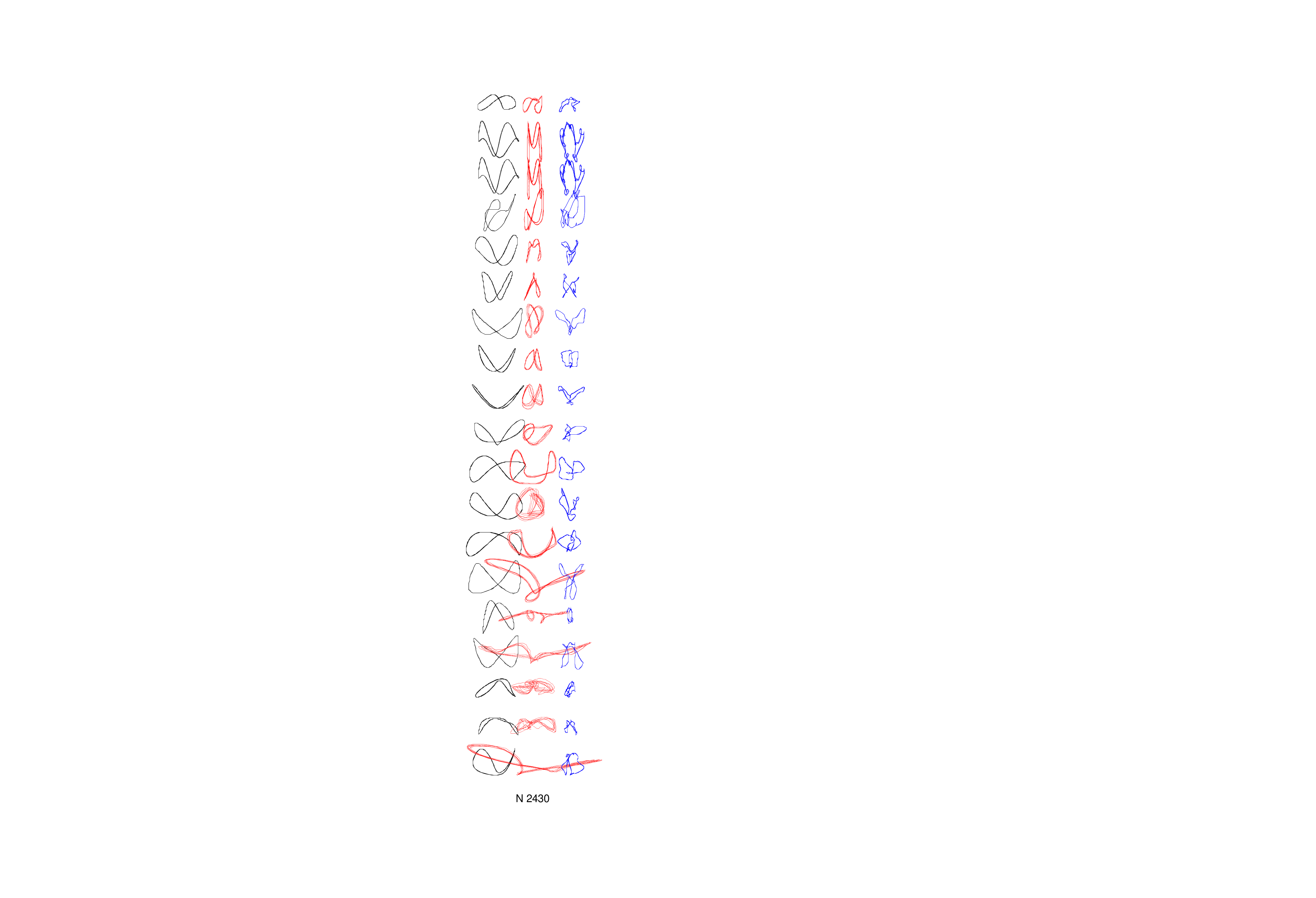}\caption{Quality of prediction on a reference data set. Cf. color code of Figure
\ref{Flo:decao_reproduction}}
\label{Flo:predict_N2430}
\end{figure}

\subsection{Dynamic analysis}

The comparison of Figure \ref{Flo:NDP2370banner} and \ref{Flo:VIVID2370banner}
is encouraging: when simulating exactly the system from which trajectories
were aquired experimentally in the first place, the simulation come
strickingly close to the experimental results. It is understood that
a simulation is good if it captures the statistical properties of
the real response. VIV being chaotic, there is no hope to reproduce
exactly any given realization of the response.

Figure \ref{Flo:VIVD2030banner} shows the result of a simulation
of a case which is also directly represented in the training data.
The simulation fails, in the sense that IL and CF vibrations are simulated
to occur at the same frequency. Two explanations are proposed:

The quality of the force prediction by the rotatron, illustrated in
Figure \ref{Flo:decao_reproduction} was declared {}``satisfactory'',
but this constitutes only an observation that the fitting procedure
is operating. On what criteria should one judge that the fit is adequate?
When training was carried out, the average norm of the difference
between training force and predicted force was found to be about 30\%
of the average norm. For a control group composed of data points from
the same experimental database, but not used in training the perceptron,
the same ratio was about 40\%. These number are high, and can be reduced
to some extend by increasing the number of hidden layers and of training
iterations, but this was not found to yield better simulations. Further
specialising the perceptron (by training it with data from test 2030)
did not lead to successful simulations. One study which might help
to understand the observations would be to find the corrective forces
that need to be added to the forces predicted by the model, to force
the simulation to track the motions observed during test 2030. This
can be achieved using dynamic inverse FEM analysis \cite{maincon04a,maincon04b,maincon08}.

Even if the above corrective forces were strictly zero, so that the
model was accurately predicting forces for the riser motion from test
2030, this would not be sufficient to ensure that the simulation adequately
mimics test 2030. A given trajectory could still have very different
stability properties in the physical system and in the simulation.
It could be that in the physical system, the trajectories follows
the {}``bottom of a valley'' while model renders it as the {}``crest
of a mountain''. A measure of stability for this is the Lyapunov
exponent \cite{lyapunov92,ott02}. Procedures exist to compute Lyapunov
exponents from experimental data, and this should be compared to Lyapunov
exponents for the simulation.

\subsection{Influence of $t_{w}$}

One issue that was explored was the adequacy of $t_{w}$: The rotatron
was trained with $t_{w}=5\cdot10^{-5}$. This corresponds roughly
to 1/4 of a cross-flow oscillation period in reduced scale for test
2370, and to a smaller fraction for other tests at lower current velocity.
It could be argued that this fraction becoming too small could be
the cause of the failure of analyzes at lower velocities (Figure \ref{Flo:VIVD2030banner}).
The rotatron was hence trained again using a suitably increased value
of $t_{w}$ . This was done twice, once with the full training set
and increase the number of Laguerre polynomials to $n=60$, and once
with only the experimental data from lower currents, and an unchanged
number of polynomials ($n=30$). Neither rotatrons allowed to perform
a successful simulation for the lower current velocities.

\subsection{Tension}

In Figure \ref{Flo:NDP2370banner}, one can observe modulations of
the positions of the vibration nodes (in particular on the cross-flow
graph). One possible explanation for this would be that in the physical
system, the tension is modulated by the vibration. This effect, if
present, is not captured by the numerical model, which assumes constant
tension. The method presented here is designed for use in a non-linear
analysis. However in this research, time was saved by using a simpler
linear structural model.

\section{Conclusions\label{sec:Conclusions}}

A model for the prediction of VIV forces given the history of velocity
of a cylindrical cross section relative to the undisturbed fluid,
has been developed. The model is closely relatied to Wiener-Laguerre
filters: the recent history of velocity is represented by the coefficients
of a Laguerre polynomial series. These coeffcients are then used to
enter a memory-less non-linear interpolation function, in this case,
a custom made neural network in which some relevant symmetry properties
were {}``hard-wired''. The neural network was trained by using forces
and displacements obtained in irregular forced motion tests on a short
cylinder.

The proposed model operates in the time domain, making it well suited
for integration into fully non-linear analyses with unsteady currents.
It further deals with in-linea and cross-flow vibrations as one inseparable
issue, which is arguably a necessity to improve on existing VIV models.

The model could provide a {}``good'' reproduction of the forces
in the training test, as well a {}``good'' prediction of forces
for {}``comparable'' trajectories. Were the model was queried with
trajectories very different from those present in the training data,
the model gave very poor results - as can be expected. Due to the
limited amount of experimental data available, the present model remains
quite specialized to a limited number of situations. What remains
unknown at this stage is the size of the training set, and of the
neural network model, necessary to create a model with some pretention
of generality.

In some in dynamic analyses of laboratory tests (NDP TN2030 and TN2430)
with a long flexible riser, the numerical solution fell into an unphysical
mode of vibration with the same frequency for in-line and cross flow
vibration. On the other hand, for another case (TN2470), some unusually
fine details were captured by the numerical model. 

From this, it is concluded that the concept has merit and deserves
to be pursued, acknowledging that more work is needed to arrive to
a pratical engineering tool.

\section{Acknowledgments}

The author is indebted to Ida Aglen, Celeste Barnardo, Trygve Kristiansen,
Carl Martin Larsen, Halvor Lie, Elizabeth Passano, Thomas Sauder,
Wu Jie and Yin Decao for inspiring discussions on the subject of VIV,
that either led to or helped the present research and writing. Thanks
are extended to CeSOS (Centre of Excellence for Ships and Offshore
Structures) at NTNU (Norwegian University of Science and Technology)
and to MARINTEK for sponsoring this research. Very special thanks
to Professor Carl Martin Larsen of CeSOS who, faced with a barrage
of dubious ideas from the author, responded by making this research
possible. 

\appendix

\section{Conventions for indexed notations\label{sec:Conventions-for-indexed}}

In the present work, index notations inspired from tensor analysis
are used. However, the present setting differs from tensor analysis
in at least three ways: 

First, we assume that we are only operating in Euclidian spaces (an
not in more general Riemannian manifolds) so that orthogonal bases
can be used. This makes it unnecessary to distinguish between co-
and contravariant bases and coordinates. Hence, only lowered indexes
appear in the present work. Incidentally, it was here assumed that
the state of the model is a point in a vector space, which is not
true when finite rotations are present and Riemanian geometry should
be introduced instead.

Second, in tensor notations, each index spans the dimension of the
manifold. In an expression like $\sigma_{ij}=C_{ijkl}\;\varepsilon_{kl}$
the indices range from 1 to 3. Following Einstein's convention, indices
$k$ and $l$ are summed over, and the relation is valid for any combination
of $i$ and $j$. The fact that the equation is valid at each point
within a solid is implicit in the notation. In the present work we
prepare for the manipulations of arrays in a computer, involving operations
that are repeated, for example for various locations alomg a riser.
If indexes $x$, $y$ and $z$ were introduced to note the position
to which the various tensors refer, one would tend to write $\sigma_{ijxyz}=C_{ijklxyz}\;\varepsilon_{klxyz}$
, which violates Einstein's convention, because no summation (or rather:
no integral) is implied over the positions.

Third, we introduce non-linear functions. These functions can combine
the values of the coordinates for some indices, and operate in parallel
on the coordinates for other indices.

Hence the following conventions are used:
\begin{enumerate}
\item By default, where an index appears more than once in a combination
of products and/or divisions, a summation over the index is implied.
If that index has a continuous range, then the {}``sum'' is an integration
over the range. Point 2, 3 and 4 specify exceptions to this rule.
\item Point 1 notwithstanding, if an equation is preceded by the symbol
$\forall$, followed by a list of indexes, then the listed indexes
are not summed over.
\item Point 1 notwithstanding, if within an equation, there is a combination
of products and/or divisions within which an index appears only once,
then no sum over that index is carried out in the whole equation.
(In any other situation than a simple term in the left hand side,
readability should be improved by using the symbol $\forall$ . )
\item Point 1 notwithstanding, if an index appears within an input to a
function, and the output of the function is multiplied or divided
by one or several terms that have the same index, then no sum within
the input to the function is carried out on that index.
\item \noindent If an index of an argument to a function is within brackets,
then the whole range of index values is used as input to one function
evaluation. For example, \textrm{$\sigma_{i}\left(y_{[j]k}\right)$
}refers to the evaluation a multiple locations ($k$) of a vector-valued
($i$) function of a vector ($j$). 
\item \noindent When the output of the function is shorthanded without explicitly
writing its input, then the indices of the input that are not within
bracket are added to the indices of the function. For example\textrm{
$\sigma_{i}\left(y_{[j]k}\right)$} can be shorthanded $\sigma_{ik}$. 
\item Derivatives of a function are noted with only the bracketed indices
of the input appearing under the fraction: $\frac{\partial\sigma_{i}}{\partial y_{j}}$
. To refer to the value of that derivative for input $k$, one writes
$\frac{\partial\sigma_{ik}}{\partial y_{j}}$ .
\end{enumerate}

\section{Rotatron gradients\label{sec:Rotatron-gradients}}

The derivative of the force predicted by the rotatron, with respect
to the Laguerre coefficients is needed in Section \ref{sec:Dynamic-analysis}.
With references to Equations \ref{eq:rot1} to \ref{eq:rot5} that
describe the rotatron, we can write\begin{eqnarray}
\frac{\partial\hat{f}_{in}}{\partial\dot{\tau}_{jl}} & = & \frac{\partial\hat{f}_{in}}{\partial\sigma_{k}}\frac{\partial\sigma_{ikn}}{\partial y_{j}}\frac{\partial y_{jkn}}{\partial\dot{\tau_{l}}}\\
 & = & V_{k}\frac{\partial\sigma_{ikn}}{\partial y_{j}}M_{kl}\end{eqnarray}
with\begin{multline}
\frac{\partial\sigma_{ikn}}{\partial y_{j}}=-\frac{1}{\left|y_{[j]kn}\right|^{3}\left(\left|y_{[j]kn}\right|^{\alpha_{k}}+1\right)^{2}}\times\\
\left[\vphantom{\frac{a}{b}}\alpha_{k}\, y_{ikn\,}y_{jkn}\left|y_{[j]kn}\right|^{\alpha_{k}}\right.\\
\left.+(-1)^{\delta_{ij}}y_{\neg ikn\,}y_{\neg jkn}\left(\left|y_{[j]kn}\right|^{\alpha_{k}}+1\right)\vphantom{\frac{a}{b}}\right]\end{multline}
Here index $i$ ranges over two values (for two directions orthogonal
to the cylinder), and $\neg i$ is the other direction than $i$.

The gradients of the rotatron with respect to its coefficients are
also needed in order to compute the gradient of the target function
with respect to the parameters $V_{k}$, $U_{k}$ and $M_{kl}$.\begin{eqnarray}
\frac{\partial\hat{f}_{in}}{\partial V_{l}} & = & \sigma_{iln}\\
\frac{\partial\hat{f}_{in}}{\partial M_{kl}} & = & \frac{\partial\hat{f}_{in}}{\partial\sigma_{k}}\frac{\partial\sigma_{ikn}}{\partial y_{j}}\frac{\partial y_{jkn}}{\partial M_{kl}}\\
 & = & V_{k}\frac{\partial\sigma_{ikn}}{\partial y_{j}}\dot{\tau}_{jln}\\
\frac{\partial\hat{f}_{in}}{\partial U_{l}} & = & V_{k}\frac{\partial\sigma_{ikn}}{\partial U_{l}}\end{eqnarray}
with\begin{equation}
\frac{\partial\sigma_{ikn}}{\partial U_{l}}=-\delta_{kl}\frac{e^{-U_{l}}y_{ikn}\left|y_{[j]kn}\right|^{\alpha_{k}-1}\log\left|y_{[j]kn}\right|}{\left(\left|y_{[j]kn}\right|^{\alpha_{k}}+1\right)^{2}}\end{equation}

\section{Inverse of $s5$}

The inverse of $s5$ (Equation \ref{eq:s5}), where $s5$ is of the
form \begin{equation}
s5_{ij}=\alpha T_{ij}+\beta\delta_{ij}\end{equation}
with\begin{equation}
\begin{cases}
T_{ij}=1 & \quad j\leq i\\
\phantom{T_{ij}}=0 & \quad j>i\end{cases}\end{equation}
can be verified to be lower triangular banded, with terms on diagonal
$i$ equal to\begin{eqnarray}
Q_{1} & = & \frac{1}{\alpha+\beta}\\
Q_{i} & = & -\frac{\alpha\beta^{i-2}}{\left(\alpha+\beta\right)^{i}}\quad i\in\left\{ 2\dots n\right\} \end{eqnarray}

\bibliographystyle{plain}
\addcontentsline{toc}{section}{\refname}\bibliography{ref}

\end{document}